\pdfoutput=1
\PassOptionsToPackage{unicode}{hyperref}
\PassOptionsToPackage{hyphens}{url}
\documentclass[
]{article}
\usepackage{xcolor}
\usepackage[margin=2.5cm]{geometry}
\usepackage{amsmath,amssymb}
\usepackage{iftex}
\ifPDFTeX
  \usepackage[T1]{fontenc}
  \usepackage[utf8]{inputenc}
  \usepackage{textcomp} 
\else 
  \usepackage{unicode-math} 
  \defaultfontfeatures{Scale=MatchLowercase}
  \defaultfontfeatures[\rmfamily]{Ligatures=TeX,Scale=1}
\fi
\usepackage{lmodern}
\ifPDFTeX\else
\fi
\IfFileExists{upquote.sty}{\usepackage{upquote}}{}
\IfFileExists{microtype.sty}{
  \usepackage[]{microtype}
  \UseMicrotypeSet[protrusion]{basicmath} 
}{}
\makeatletter
\@ifundefined{KOMAClassName}{
  \IfFileExists{parskip.sty}{%
    \usepackage{parskip}
  }{
    \setlength{\parindent}{0pt}
    \setlength{\parskip}{6pt plus 2pt minus 1pt}}
}{
  \KOMAoptions{parskip=half}}
\makeatother
\usepackage{longtable,booktabs,array}
\usepackage{calc} 
\usepackage{etoolbox}
\makeatletter
\patchcmd\longtable{\par}{\if@noskipsec\mbox{}\fi\par}{}{}
\makeatother
\IfFileExists{footnotehyper.sty}{\usepackage{footnotehyper}}{\usepackage{footnote}}
\makesavenoteenv{longtable}
\usepackage{graphicx}
\makeatletter
\newsavebox\pandoc@box
\newcommand*\pandocbounded[1]{
  \sbox\pandoc@box{#1}%
  \Gscale@div\@tempa{\textheight}{\dimexpr\ht\pandoc@box+\dp\pandoc@box\relax}%
  \Gscale@div\@tempb{\linewidth}{\wd\pandoc@box}%
  \ifdim\@tempb\p@<\@tempa\p@\let\@tempa\@tempb\fi
  \ifdim\@tempa\p@<\p@\scalebox{\@tempa}{\usebox\pandoc@box}%
  \else\usebox{\pandoc@box}%
  \fi%
}
\def\fps@figure{htbp}
\makeatother
\providecommand{\tightlist}{%
  \setlength{\itemsep}{0pt}\setlength{\parskip}{0pt}}
\usepackage{microtype}
\usepackage{titlesec}
\usepackage{booktabs}
\usepackage{array}
\usepackage{longtable}
\usepackage{ragged2e}
\usepackage{etoolbox}
\usepackage{caption}
\usepackage{fancyhdr}
\usepackage{setspace}
\usepackage{newunicodechar}

\newunicodechar{✓}{$\checkmark$}
\newunicodechar{∈}{$\in$}
\newunicodechar{∃}{$\exists$}
\newunicodechar{≠}{$\neq$}
\newunicodechar{≤}{$\le$}
\newunicodechar{≥}{$\ge$}
\newunicodechar{→}{$\rightarrow$}
\newunicodechar{₁}{$_1$}
\newunicodechar{₂}{$_2$}
\newunicodechar{ₖ}{$_k$}
\newunicodechar{─}{-}
\newunicodechar{└}{+}
\newunicodechar{├}{+}
\newunicodechar{│}{|}

\titleformat{\section}{\Large\bfseries}{\thesection.}{0.6em}{}
\titleformat{\subsection}{\large\bfseries}{\thesubsection.}{0.6em}{}
\titleformat{\subsubsection}{\normalsize\bfseries}{\thesubsubsection.}{0.6em}{}

\titlespacing*{\section}{0pt}{1.4em}{0.6em}
\titlespacing*{\subsection}{0pt}{1.1em}{0.4em}
\titlespacing*{\subsubsection}{0pt}{0.8em}{0.3em}

\renewcommand{\arraystretch}{1.08}
\AtBeginEnvironment{longtable}{\small}
\newcolumntype{L}[1]{>{\RaggedRight\arraybackslash}p{#1}}
\newcolumntype{C}[1]{>{\Centering\arraybackslash}p{#1}}
\newcolumntype{R}[1]{>{\RaggedLeft\arraybackslash}p{#1}}

\newcommand{\codeid}[1]{{\small\begingroup\urlstyle{tt}\nolinkurl{#1}\endgroup}}
\usepackage{pdflscape}
\usepackage{float}
\usepackage{needspace}
\usepackage{bookmark}
\IfFileExists{xurl.sty}{\usepackage{xurl}}{} 
\hypersetup{
  hidelinks,
  pdfcreator={LaTeX via pandoc}}

\author{}
\date{}

\begin{document}

\section{AppSec Core: An Ontology for Security-by-Design Software
Engineering}\label{appsec-core-an-ontology-for-security-by-design-software-engineering}

\textbf{Pedro Farinha}\\
Shiftleft - Secure Software Engineering, Lda.\\
ORCID 0009-0001-0569-9020 ·
\href{mailto:pedro.farinha@shiftleft.pt}{\nolinkurl{pedro.farinha@shiftleft.pt}}

\textbf{Keywords:} secure software engineering, security by design,
application security, security requirements, ontology, traceability.

\begin{center}\rule{0.5\linewidth}{0.5pt}\end{center}

\subsection{Abstract}\label{abstract}

When engineers or coding agents change an application, a dependency, a
configuration or a build pipeline, they need to know three things: which
security rules apply, how to meet them, and what evidence to leave.
Today the answers are spread over many standards, frameworks, laws and
lists of weaknesses, and each one uses its own words and level of
detail. This paper presents AppSec Core, a model of the security work
that engineers do. It says what must be true, which practices get there,
which mechanisms do the job and which artefacts show that it was done.
It has 259 elements in ten domains. We built it from a
security-by-design manual written from practice, and then checked it
against 31 outside sources, ten of them about AI systems and agents.
Along the way we made three changes, which added five objectives and
made one domain wider. We did not need a new domain. Most objectives are
backed by several sources: the typical objective is used by six of the
twenty technical sources, and 194 of the 202 objectives, practices and
mechanisms are linked to items in the sources. For the domain of a
change, the model gives back a closed set of typed elements, and for a
source, it gives back the objectives that the source points to. We keep
the model apart from the manual, the selection rules and the server that
delivers the result, and we follow one change, adding a third-party
package, through all four. When a code generator received the
requirements, the share of security tests passed on a public benchmark
went from 77.4\% to 93.0\%. On a second benchmark, the result was close
to that of the benchmark's own ``oracle'' prompt, which is written with
the tests in mind.

\begin{center}\rule{0.5\linewidth}{0.5pt}\end{center}

\subsection{1. Introduction}\label{introduction}

\subsubsection{1.1 The problem at a point of
change}\label{the-problem-at-a-point-of-change}

Think of an engineer who adds a third-party package to a web service,
changes the authentication middleware, or edits the pipeline that builds
and releases the service. Each of these changes raises the same three
questions. Which security rules apply here? How can the team meet them?
And what evidence should the change leave, so that a reviewer or an
auditor can see that they were met?

The answers exist, but no single document holds them. Development
frameworks such as the Secure Software Development Framework (SSDF) of
the US National Institute of Standards and Technology (NIST) {[}1{]}
describe practices across the lifecycle. Verification standards such as
the Application Security Verification Standard (ASVS) of the Open
Worldwide Application Security Project (OWASP) {[}2{]} list things about
the application that can be tested. Supply-chain models such as
Supply-chain Levels for Software Artifacts (SLSA) {[}3{]} deal with how
builds are protected, and control lists such as the Critical Security
Controls of the Center for Internet Security (CIS) {[}4{]} deal with the
systems of the whole organisation. Lists of weaknesses and attacks, such
as the Common Weakness Enumeration (CWE) {[}5{]} and the Common Attack
Pattern Enumeration and Classification (CAPEC) {[}6{]}, describe what
can go wrong rather than what to do. On top of these, laws such as the
EU Cyber Resilience Act (CRA) {[}7{]}, the NIS2 Directive {[}8{]} and
the Digital Operational Resilience Act (DORA) {[}9{]} add duties written
in legal language.

These sources overlap a lot, but they do not cut the work into the same
pieces or at the same level of detail. One ASVS requirement can match
part of an SSDF task, a CIS safeguard and the fix for several CWE
weaknesses. So an engineer who wants a full answer for one change has to
read several sources and work out how they fit together, and then do it
again for the next change and the next project. When this does not
happen, compliance is checked later, in testing or in an audit.

Coding agents now have the same problem. An agent that writes or changes
code usually gets a functional task, and the security rules for that
task do not reach it unless someone writes them down. In an earlier
study we gave the code generator the requirements that applied, at the
moment it wrote the code, and measured the effect on two public
benchmarks {[}10{]}. That study needed one source that could say which
requirements apply to a given task. This paper describes how we built
that source.

\subsubsection{1.2 Where AppSec Core comes
from}\label{where-appsec-core-comes-from}

The work started from a manual. The SbD-ToE manual {[}11{]} describes a
practical way to do application security across the software lifecycle:
who does what, in which phase, at which risk level, and how the work is
checked. A manual like this is one choice of method, and other
organisations will split roles and phases in other ways. But what the
rules ask for should not depend on that choice, because the same
standards and laws apply whatever the method. We wanted to know whether
what the manual asks teams to do covers what the outside sources ask
for. We first tried to compare the manual with each source in turn, and
it did not work, because each source uses its own units and level of
detail. So we had to write the engineering content down once, in one
common form, and compare both the manual and every source with that.

AppSec Core is that common reference. It is a model of the security work
of engineers, written from practice, and not just a list of requirements
collected from the sources. Each element also records the source items
that ask for it. This is what lets the manual say: apply this user story
or this requirement, and you meet this objective, which these outside
sources require. The chain from a user story to the sources is not yet
complete for every chapter of the manual (Section 6), but this is what
the model was designed for.

The idea behind it is not new. Secure software engineering has a limited
set of things to get right, and the many publications in the field keep
coming back to them, in different words and sometimes in the language of
one business or one regulator. We use the mappings in this paper to test
that idea for AppSec Core: they show how much of each source the model
covers, and whether the sources support the way the model was built.

\subsubsection{1.3 Research question and
scope}\label{research-question-and-scope}

The paper asks one question.

\begin{quote}
\textbf{RQ.} To what extent can a bounded model of engineering work hold
the relevant content of many different application-security sources,
keep their traceability, and help decide which security rules apply to a
concrete change?
\end{quote}

Inside its boundary, AppSec Core gives a working answer to the question
``what is application security?''. The answer is a set of conditions
that must be true for the software and for the process that builds it
(control objectives), the ways a team can meet them (practices and
mechanisms), and the evidence that shows they were met (artefacts). This
working definition has one purpose: to tell an engineer or a coding
agent, for a given change, which conditions must be true, how to meet
them and what evidence shows it, and to trace each of these back to the
sources that ask for it. We judge the model by how well it does this
job, not by formal completeness. So it has to be broad enough to express
what the sources ask for, stable enough that new sources do not force a
redesign, and precise enough that the same declared context always gives
the same expectations. The model does not claim to be the one true
definition of application security, and meeting every objective that
applies does not by itself make an application secure.

\subsubsection{1.4 Contributions}\label{contributions}

The paper makes three contributions.

\begin{itemize}
\tightlist
\item
  \textbf{The engineering model.} AppSec Core v1 describes application
  security as control objectives, practices, mechanisms and artefacts,
  with fixed relations between them. The elements sit in ten domains
  inside a declared boundary and have stable identifiers, and the
  objectives, practices and mechanisms link to the source items that ask
  for them (Section 2).
\item
  \textbf{How we built and checked it.} We built the model from practice
  and then checked it against 31 outside sources. We describe how the
  sources were broken down and mapped, what we did with content outside
  the boundary, the rule under which the model could change, and what
  all this tells us about how well the model covers the sources, whether
  the mappings are right, and how the model is used (Sections 3, 4, 5
  and 8).
\item
  \textbf{How it is used in practice.} We separate the jobs of the
  model, the manual, the selection rules and the delivery server, as
  they are built today, and we follow one change through all four, from
  the declared context to the expected evidence and the sources behind
  it (Sections 6 and 7).
\end{itemize}

\subsubsection{1.5 Relation to earlier
work}\label{relation-to-earlier-work}

This paper is the main presentation of AppSec Core. It brings together
results from separate preprints, and it replaces the preprint ``AppSec
Core: A Normalized Ontology for Security Requirements Across
Heterogeneous Frameworks'' {[}12{]}, which described version 0 of the
model. Three other preprints remain the detailed references. The formal
artefact, with its Web Ontology Language (OWL 2) export, its Shapes
Constraint Language (SHACL) constraints and its validation, is described
in {[}13{]}. The design cycle that checked the model against the
sources, with the similarity pipeline and the round on sources about
artificial intelligence (AI), is described in {[}14{]}. How the manual
and a knowledge graph are built from the model is described in {[}15{]}.
This paper reuses the counts of the formal artefact and the sources of
the design cycle, and where those preprints said something that the
current data do not support, we correct it at that point. The study of
how delivered requirements affect generated code {[}10{]} is cited for
its results and not repeated here. The workshop paper {[}16{]} shows how
the model is used at a point of change, and we use it as evidence of use
in Section 8.3.

\begin{center}\rule{0.5\linewidth}{0.5pt}\end{center}

\subsection{2. The engineering model}\label{the-engineering-model}

\subsubsection{2.1 Element types and
relations}\label{element-types-and-relations}

AppSec Core has four main kinds of element, and each one answers a
question that an engineer asks about a change (Figure 1).

\begin{itemize}
\tightlist
\item
  A \textbf{control objective} says what must be true, in a form that
  can be checked. An example is ACO-SCBI-001, \emph{Dependency Inventory
  and SBOM Traceability}, where SBOM stands for software bill of
  materials.
\item
  A \textbf{practice} says how a team reaches one or more objectives,
  for example ACP-SCBI-001, \emph{Build-Linked SBOM Generation}.
\item
  A \textbf{mechanism} says what is used to put an objective in place,
  such as a tool, a configuration or a runtime control. An example is
  ACM-SCBI-005, \emph{Build and Image Inventory Generation}.
\item
  An \textbf{artefact} is an output that shows that an objective was
  met, such as an SBOM, a scan report or an approval record.
\end{itemize}

\begin{figure}
\centering
\includegraphics[width=1\linewidth,height=\textheight,keepaspectratio,alt={AppSec Core at a glance. Each domain holds control objectives, the practices that realise them, the mechanisms that implement them and the artefacts they expect. Source items reach the model through the curated mappings, and manual requirements through the manual controls. Selection picks manual requirements and serves their whole domain. Dashed lines: evidence patterns are declared in the ontology and kept in the manual, where they name artefacts.}]{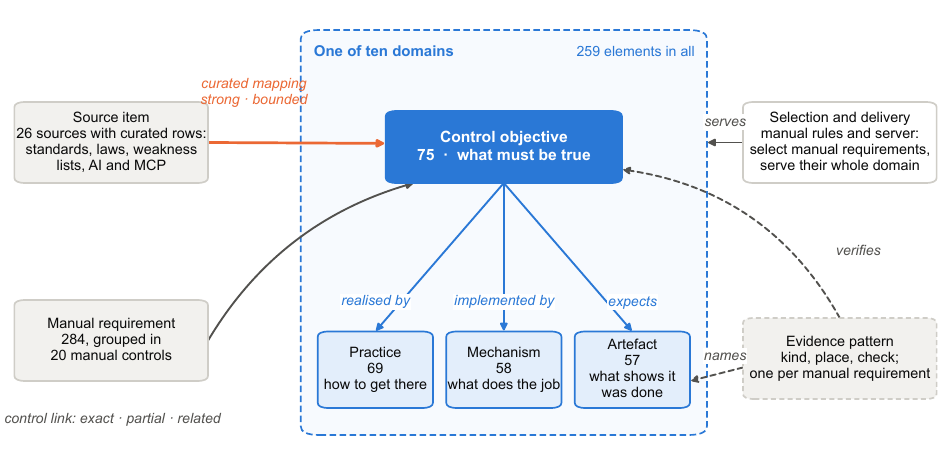}
\caption{AppSec Core at a glance. Each domain holds control objectives,
the practices that realise them, the mechanisms that implement them and
the artefacts they expect. Source items reach the model through the
curated mappings, and manual requirements through the manual controls.
Selection picks manual requirements and serves their whole domain.
Dashed lines: evidence patterns are declared in the ontology and kept in
the manual, where they name artefacts.}
\end{figure}

Each control objective is linked to at least one practice that reaches
it and at least one mechanism that puts it in place, and it can be
linked to the artefacts it expects. Practices and mechanisms are not
linked to each other directly, so every chain goes through an objective.
This keeps the first question, what must be true, at the centre of every
use of the model. The released version has 75 objectives, 69 practices,
58 mechanisms and 57 artefacts.

\par\noindent\begin{minipage}{\linewidth}

\textbf{Box 1.} One control objective in full, as written in the domain
contracts of the public release {[}17{]}.

\begin{quote}
\textbf{ACO-IAT-006} \emph{Access Abuse Detection and Auditability}
(identity, access and session trust domain)\\
\textbf{Statement.} Access surfaces must emit sufficient telemetry to
detect abuse, prove rejected access and audit externally visible calls
with reviewable context.\\
\textbf{Expected outcome.} Rate limits, access denials and external-call
activity are visible enough to support operational review and post-event
investigation.\\
\textbf{Practice.} ACP-IAT-006 \emph{Access Abuse Monitoring and Audit
Trail}.\\
\textbf{Mechanisms.} ACM-IAT-003 \emph{Periodic Review and Access
Audit}; ACM-IAT-006 \emph{Structured Logging and Effective Configuration
Recording}.\\
\textbf{Expected artefacts.} ACA-IAT-003 \emph{Log Stream}, ACA-IAT-004
\emph{Policy Report} and ACA-IAT-005 \emph{Alert Policy}.
\end{quote}

\end{minipage}\par\medskip

A fifth kind of element, the evidence pattern, describes the evidence
expected for a requirement: what kind of evidence it is, where to find
it, how to check it and which artefacts it should produce. Today
evidence patterns live in the manual, one for each manual requirement,
and they reach AppSec Core through the domain of their requirement and
through the artefacts they name. The ontology declares the type, but it
is not yet filled in with AppSec Core elements. Evidence patterns are
also not mapped to the outside sources, because those mappings stop at
objectives, practices and mechanisms {[}13{]}.

\subsubsection{2.2 Domains and boundary}\label{domains-and-boundary}

The elements are grouped in ten domains. Each domain has a written
contract that says what it covers, what it leaves out and where in the
lifecycle its controls apply. Table 1 lists the domains.

\textbf{Table 1.} The ten domains of AppSec Core v1 and their number of
elements (objectives, practices, mechanisms and artefacts).

\begin{longtable}[]{@{}
  >{\raggedright\arraybackslash}p{(\linewidth - 6\tabcolsep) * \real{0.2308}}
  >{\raggedright\arraybackslash}p{(\linewidth - 6\tabcolsep) * \real{0.2308}}
  >{\raggedright\arraybackslash}p{(\linewidth - 6\tabcolsep) * \real{0.2308}}
  >{\raggedleft\arraybackslash}p{(\linewidth - 6\tabcolsep) * \real{0.3077}}@{}}
\toprule\noalign{}
\begin{minipage}[b]{\linewidth}\raggedright
Domain
\end{minipage} & \begin{minipage}[b]{\linewidth}\raggedright
Prefix
\end{minipage} & \begin{minipage}[b]{\linewidth}\raggedright
Scope
\end{minipage} & \begin{minipage}[b]{\linewidth}\raggedleft
Elements
\end{minipage} \\
\midrule\noalign{}
\endhead
\bottomrule\noalign{}
\endlastfoot
Supply chain and build integrity & SCBI & Dependencies, build
provenance, artefact signing & 30 \\
Identity, access and session trust & IAT & Authentication,
authorisation, sessions & 25 \\
Architecture and trust boundaries & ATB & Boundaries, exposure,
isolation & 23 \\
Testing and security validation & TSV & Security testing and empirical
assurance & 25 \\
Threat modelling and risk disposition & TMR & Threat models, risk
decisions, requirement traceability & 29 \\
Secrets and protected configuration & SPC & Secret handling, protected
configuration, operational identities & 19 \\
Input and output data safety & IVF & Validation, safe parsing, output
encoding, controlled failure & 26 \\
Integration trust & ITS & Service-to-service and third-party integration
& 23 \\
Release and rollout & RPR & Promotion, controlled rollout, rollback
readiness & 39 \\
Security logging and audit & SLG & Security events, audit trail, central
logging & 20 \\
\textbf{Total} & & & \textbf{259} \\
\end{longtable}

Every element belongs to exactly one domain, the one where its control
is applied. Relations stay inside a domain: an objective, its practices,
its mechanisms and its artefacts are always in the same domain.

The boundary of the model is the engineering work done while building
and maintaining an application. Content where the main actor is the
organisation and not the engineering team is outside it, for example
board reporting, contracts with suppliers, procedures for notifying
regulators and staff training. Physical security and general IT
operations are outside it too, and so is memory safety, which we leave
to languages and toolchains. Cryptography as a field is left out, but
the way applications use cryptography is in. The boundary keeps the
model small enough to use for a single change. Content outside the
boundary is not lost: it is recorded as out of scope with a reason, and
where it still concerns application security, the manual covers it in
its text (Section 3).

\subsubsection{2.3 Identifiers, traceability and
evolution}\label{identifiers-traceability-and-evolution}

Every element has a stable identifier of the form
\texttt{AC\textless{}T\textgreater{}-\textless{}DOMAIN\textgreater{}-\textless{}NNN\textgreater{}},
where \texttt{\textless{}T\textgreater{}} is O, P, M or A for the type
of element. Identifiers are never reused, and every identifier of
version 0 still exists in version 1. Each objective, practice and
mechanism can be linked to the items of outside sources that ask for it.
A link records a primary objective, optional secondary objectives, the
practice and mechanisms that apply, a strength and a short reason
(Section 4.4). These links record where the sources cover the model. The
model itself was defined from the manual (Section 3), and the links are
not meant as a crosswalk between the sources. Artefacts are reached
through the objectives that expect them. The model changes only under a
set rule, described in Section 5, and its version number follows the
kind of change.

\subsubsection{2.4 Why an ontology}\label{why-an-ontology}

We publish the model as an ontology because, to use it, we need a typed,
closed structure that a machine can check. An engineer or an agent who
asks for the objectives of a domain must get the same objectives,
practices, mechanisms and artefacts every time, and a tool must be able
to reject a graph where an objective has no practice or an element sits
in two domains. The OWL 2 export {[}18{]} declares the types, the
domains and the relations, and the SHACL constraints {[}19{]} check
these rules on the data graph, as closed-world validation. The namespace
of the ontology is
\texttt{https://securitybydesign.dev/ontology/appsec-core/v1\#}. The
released graph passes validation with pyshacl 0.31.0 with no violations,
and the OOPS! pitfall scanner {[}20{]} reports no critical or important
pitfalls {[}13{]}. Other catalogues, such as OpenCRE, also have stable
identifiers and can be used by programs. What the ontology adds here is
the typed structure and its constraints (Section 9). Whether each
mapping is right is a separate question (Section 8.2).

\begin{center}\rule{0.5\linewidth}{0.5pt}\end{center}

\subsection{3. Built from practice}\label{built-from-practice}

The manual came first, and AppSec Core was taken from it. The author
wrote the manual over several years of consulting work. It uses the
published guidance, but it is organised around the work of engineers and
not around any one source {[}11{]}. We got the domains and element types
of the model by reading the manual and pulling out the structure that
keeps coming back in its requirements: objectives, practices, mechanisms
and artefacts {[}12{]}. In doing so, the model kept the ``what'' of the
manual and left out the ``who'' and the ``when'', which belong to the
method.

We kept the model small on purpose, so that it would not grow just by
piling up content. We fixed the ten domains first. Together they set the
boundary of secure software engineering as we use the term here: the
areas an engineer has to keep in mind while building and maintaining an
application. Inside each domain we then aimed for a small, fixed number
of control objectives: seven, with six specific ones and one composite
that groups them. This number was our choice of method, not a property
of the domain {[}12{]}. Its job was to push us towards objectives that
do not overlap and that together cover the domain, so that each
practice, mechanism and artefact has one clear objective to attach to.
We adjusted the objectives in this first pass until the set looked
complete enough. Version 0 ended with exactly seven objectives in each
of the ten domains: 70 objectives, 63 practices, 48 mechanisms and 53
artefacts, 234 elements in all. After that, the number could change only
through the change rule. Three domains now have more than seven
objectives. Release and rollout has ten after the AppSec Core Change
Request ACR-001, and threat modelling and input and output data safety
have eight each, after ACR-002 and ACR-004 (Section 5).

A model taken from the manual cannot show on its own that the manual is
complete, because it would only repeat the manual. Its value as a
reference comes from the next step, where we checked it against outside
sources that the author did not write (Section 4) and changed it when
those sources gave good reason (Section 5). After that, the manual was
aligned again with the model and checked against the same sources, and
the gaps we found were closed in the manual {[}15{]}. So the model is
the place where the manual and the outside sources meet, and neither is
taken as correct without that comparison. How much the manual covers is
not measured in this paper.

\begin{center}\rule{0.5\linewidth}{0.5pt}\end{center}

\subsection{4. Checking the model against outside
sources}\label{checking-the-model-against-outside-sources}

\subsubsection{4.1 The corpus}\label{the-corpus}

The check had two rounds. The first round asked whether the boundary of
the model was right: whether the model covered enough of application
security across its whole width, or whether something that the sources
treat as application security was missing and should be brought in. For
this we picked five sources that look at the field from different
angles: a lifecycle framework (SSDF {[}1{]}), a verification standard
(ASVS {[}2{]}), a supply-chain model (SLSA {[}3{]}), a list of controls
(CIS Controls {[}4{]}) and a list of attack patterns (CAPEC {[}6{]}).
The round gave two results. It found two possible gaps, which we carried
into the second round and later admitted (ACR-001 and ACR-002, Section
5.3). It also made us decide when the sources are a good enough reason
to change AppSec Core, and we set the thresholds of the change rule from
what we saw in this round (Section 5.1). Five sources are too few to
know whether the boundary would hold for a wider set {[}14{]}, so the
second round extended the check to 31 sources.

In the second round we chose sources from different publishers and with
different views of the field, without favouring sources that would fit
the model. They fall into three groups that we handled differently, and
we keep the groups apart in the rest of the paper.

\begin{itemize}
\tightlist
\item
  \textbf{Group A, twenty technical sources mapped row by row.} ASVS
  5.0.0 {[}2{]}, CAPEC 3.9 {[}6{]}, CIS Controls 8.1.2 {[}4{]}, CWE
  4.19.1 (software development view) {[}5{]}, NIST SP 800-53 Rev.~5
  {[}21{]}, the OWASP Software Assurance Maturity Model (SAMM) 2.1
  {[}22{]}, the OWASP DevSecOps Maturity Model (DSOMM) {[}23{]}, OWASP
  Top 10:2021 {[}24{]}, OWASP Proactive Controls 2018 {[}25{]}, the
  Payment Card Industry (PCI) Secure Software Lifecycle (Secure SLC)
  standard 1.1 {[}26{]}, three SAFECode guides {[}27, 28, 29{]}, SLSA
  1.0 Build Track {[}3{]}, SSDF 1.1 {[}1{]}, four documents on the
  security of the Model Context Protocol (MCP) {[}30, 31, 32, 33{]}, and
  the framework of good cybersecurity practices for AI of the European
  Union Agency for Cybersecurity (ENISA) {[}34{]}.
\item
  \textbf{Group B, six regulations and industry standards mapped with a
  heuristic.} The CRA {[}7{]}, DORA {[}9{]}, NIS2 {[}8{]}, the General
  Data Protection Regulation (GDPR) {[}35{]}, the PCI Data Security
  Standard (DSS) 4.0.1 {[}36{]} and the Security Rule of the US Health
  Insurance Portability and Accountability Act (HIPAA) {[}37{]}.
\item
  \textbf{Group C, five more sources on AI and machine-learning systems,
  compared by text similarity, with a review of the items left without a
  link.} MITRE ATLAS {[}38{]}, the OWASP Top 10 for Large Language Model
  (LLM) Applications 2025 {[}39{]}, the OWASP Machine Learning Security
  Top 10 {[}40{]}, NIST AI 100-2 E2025 {[}41{]} and the NIST AI Risk
  Management Framework {[}42{]}. We added these last, on purpose, to see
  whether recent material on AI systems would need a new structure.
\end{itemize}

Ten of the 31 sources are about AI systems or AI agents: the four MCP
documents and the ENISA framework in group A, and the five sources of
group C. Some sources have had new versions since we mapped them, for
example SLSA, OWASP SAMM, the OWASP Top 10, CWE and NIST SP 800-53. The
mappings refer to the versions listed above.

\subsubsection{4.2 Two layers of evidence}\label{two-layers-of-evidence}

The check produced two layers of evidence, and they should be read in
different ways.

The \textbf{curated mappings} are files with one row per source item,
for the 26 sources of groups A and B. Each mapped row records a primary
objective, optional secondary objectives, the practice and mechanisms, a
strength and a reason. We call the objectives that a row is mapped to
its anchors, primary or secondary. These files are the basis for the
traceability of the model and for the counts in Section 8.

The \textbf{similarity substrate} is a separate layer, computed in a
fixed way over all 31 sources {[}14{]}. The sources were broken down
into 4,935 rows in all, grouped into 3,861 source items, and a
sentence-embedding model (all-MiniLM-L6-v2 {[}43{]}) linked a row to an
objective, practice or mechanism whenever their cosine similarity passed
a threshold calibrated for each kind of element (about 0.41). This gave
18,673 links, which the substrate calls claims. Items with no link above
the threshold were put aside for review. We used the substrate to look
for content that the model could not hold, for example groups of items
that link only weakly to every element, and for the sources of group C
it was the starting point of the review described in Section 8.1. Its
links are only candidates, to be reviewed. In an internal spot check by
the author, with 30 links from each layer, 63\% of the substrate's links
to objectives were judged wrong, against 17\% for the curated mappings.
This is why no count in this paper uses the substrate as a statement of
what a source asks for.

Table 2 shows how the counts in this paper relate to those in the
earlier preprints, because the same set of sources has been counted in
different units.

\textbf{Table 2.} Counting units used for the corpus.

\begin{longtable}[]{@{}
  >{\raggedleft\arraybackslash}p{(\linewidth - 4\tabcolsep) * \real{0.4000}}
  >{\raggedright\arraybackslash}p{(\linewidth - 4\tabcolsep) * \real{0.3000}}
  >{\raggedright\arraybackslash}p{(\linewidth - 4\tabcolsep) * \real{0.3000}}@{}}
\toprule\noalign{}
\begin{minipage}[b]{\linewidth}\raggedleft
Count
\end{minipage} & \begin{minipage}[b]{\linewidth}\raggedright
Unit
\end{minipage} & \begin{minipage}[b]{\linewidth}\raggedright
Scope
\end{minipage} \\
\midrule\noalign{}
\endhead
\bottomrule\noalign{}
\endlastfoot
4,935 & rows of the similarity substrate after decomposition & 31
sources \\
3,861 & source items of the similarity substrate & 31 sources \\
18,673 & similarity links (claims) from rows to elements & 31 sources \\
3,671 & rows of the curated mappings & 26 sources (groups A and B) \\
3,389 & rows of the curated mappings & 20 sources (group A) \\
\end{longtable}

For the same 26 sources there are 3,671 curated rows but 3,457 substrate
items. The difference comes from rows that the curated files keep and
the substrate did not break down: mostly headings and context rows of
ASVS and SAMM, the cross-reference tasks of SSDF and duplicate
identifiers in the CIS file. An earlier count of 3,957 rows {[}13{]}
also included an older version of ASVS that is not part of this set.

\subsubsection{4.3 How the curated mappings were
made}\label{how-the-curated-mappings-were-made}

The curator of the mappings is the author. For each source in group A, a
script applied rules written for the structure of that source, and the
curator then reviewed the result. The rules used the section structure
and the vocabulary of each source. For example, SSDF tasks take the
mapping of their practice unless a task needs its own, CAPEC attack
patterns inherit through the CAPEC hierarchy from 61 top-level patterns
that were mapped by hand, and CWE weaknesses are mapped by theme rules
from their description and their listed mitigations. About half of the
mapped rows were inherited from a parent or a section and not decided
one by one. The curator then reviewed the output against the source
text.

The curated mappings went through two recorded states. \textbf{Mapping
state 1} (13--22 April 2026) created the files against version 0, moved
the content of ACR-001 and ACR-002 to the new elements, and applied 32
recorded corrections after a quality review. A language model, acting as
a reviewer, suggested some of these corrections, and the curator
accepted them. \textbf{Mapping state 2} (17--19 October 2026) came after
an audit of how each source was covered. The audit found three problems
that came up again and again: mappings inherited from a parent without a
review of each item, provisional anchors that nobody had looked at again
after ACR-001, and the heuristic mapping of the regulations and industry
standards in group B. Four corrections were made in this state: one SSDF
task was moved to a better objective, 31 items were moved to the
objective added by ACR-004 (Section 5.4), the SLSA anchors were reviewed
against the text of the specification, and so were the SSDF anchors. An
AI agent made these edits, and the author approved them. The first two
problems were fixed only where they were found, and the third is still
there, as described below. All counts in this paper refer to mapping
state 2.

The six sources of group B were mapped with one recorded heuristic.
Selected articles of each regulation, and the requirements of PCI DSS,
are first assigned to a domain and then to an objective in that domain.
All rows mapped this way have the lowest strength, and their records say
how they were made, so that later uses can give them less weight. The
sources of group C have no curated mappings.

The curated mappings were not rebuilt from scratch after each change to
the model, and every identifier in mapping state 2 exists in version 1.
The similarity substrate, on the other hand, was computed again after
the last change request, and when the AI sources were added, the 3,457
items of the 26 earlier sources gave exactly the same links {[}14{]}.

\subsubsection{4.4 Mapping strength and what a link
means}\label{mapping-strength-and-what-a-link-means}

A mapped row records how well the item fits the element. A \emph{strong}
link means that the item says what the objective says. A \emph{bounded}
link means that it says part of it, or says it for a narrower or wider
scope. Two of the scripts also wrote \emph{direct}, which in practice
means the same as strong. In the twenty sources of group A there are
1,910 strong or direct rows and 546 bounded rows, 789 rows are recorded
as outside the boundary, 142 are headings or context, and 2 are
cross-references.

A link does not always mean the same thing, because the sources support
the model in different ways. A verification requirement in ASVS or a
task in SSDF says what to do, and it is mapped to the objective it
states. A weakness in CWE or an attack pattern in CAPEC says what can go
wrong, and it is mapped to the objective whose control prevents it, so
the link reads the source the other way round. An article of a law
states a legal duty, often as outcomes and organisational measures, and
only its engineering part can be mapped, through the domain it concerns.
So a link says that the objective covers, mitigates or contributes to
the source item, depending on the kind of source, with the strength
recorded. It does not say that the two mean exactly the same thing, and
meeting a linked objective does not prove compliance with the source.

\subsubsection{4.5 Content outside the
boundary}\label{content-outside-the-boundary}

Not every item belongs in the model, and the check had to record what
was left out instead of ignoring it. In the 26 curated sources, 814 rows
were recorded as outside the boundary and 142 as headings or context.
CWE has the largest share, with 256 of its 400 rows outside the
boundary, mostly weaknesses about general code quality, human factors,
memory safety and cryptographic algorithms. In NIST SP 800-53, 327 of
1,196 controls were outside, mostly controls of the organisation, the
physical environment and staff. The design of the cycle had two places
for excluded content: the manual, for content that concerns application
security at the level of the organisation, and plain exclusion, for
content that does not concern application security. The curated files
record the exclusion and its reason, but not this split for each row, so
we cannot say how many excluded rows went to each place.

\begin{center}\rule{0.5\linewidth}{0.5pt}\end{center}

\subsection{5. How the model changes}\label{how-the-model-changes}

\subsubsection{5.1 The change rule}\label{the-change-rule}

We set the rule after the first round, when its two possible gaps raised
the question of when the sources are a good enough reason for a change.
Any check against outside sources will find items that fit badly. If
every such item could change the model, the model would grow with each
new source and would stop being a small, stable set. If no item could
change it, the check would only measure fit and would never improve the
model. A change rule, the AppSec Core Change Request (ACR), decides when
the sources give enough reason for a change.

The rule gives criteria, and the curator decides. A candidate change is
normally considered when three conditions are met {[}13, 14{]}. At least
three independent sources point to the same gap. The content is
engineering practice, because organisational and legal content with no
engineering control belongs in the text of the manual. And the candidate
is different from the elements that already exist, not the same thing in
other words. A candidate that meets these three conditions is then
reviewed at three levels. The review asks whether the manual already
covers the content in its text, whether the content is already spread
over existing practices and only needs better traceability, and whether
there is a real gap at the level of an objective or a domain. For
candidates about AI agents and their tools, the review also looks at how
serious the exposure is, for example whether it sits in the agent that
calls the tool or in the tool itself. New technology alone is not a
reason for a change. In our experience, earlier waves such as the web,
mobile, cloud and containers were absorbed by extending practices and
mechanisms, not by new engineering categories {[}14{]}. A candidate that
fails any of these questions is not admitted, and a refusal is recorded
in the same way as an admission. When the curator makes an exception to
the criteria, the record says why.

The result of a review falls into one of three tiers. In tier A, the
default, the answer stays in the text of the manual or in an extension
of an existing practice, and the objectives do not change. Tier B adds
elements, such as a new objective with its practices and mechanisms, and
needs at least three independent sources. Tier C adds a domain and needs
at least four sources and evidence that the existing domains cannot hold
the content. A separate maintenance rule allows a mechanism to be added
inside a domain when it gives more detail on how an existing objective
is put in place, as long as it stays inside the domain contract and is
recorded with its identifier. If there are more than five such additions
in one domain, or twenty in total between major versions, the case goes
to a change request; this threshold is still provisional {[}13{]}. The
version number follows these cases: a new domain or relation changes the
major version, a tier B change the minor version, and a maintenance
addition the patch version. Version 1 is the first release we tagged as
stable, once no change request was open; earlier states, including the
version described in {[}12{]}, were version 0. The major number marks
the shape of the model, its ten domains and its structure, and we keep a
new major version for a change of that shape.

\subsubsection{5.2 Independence of
sources}\label{independence-of-sources}

The rule asks for independent sources, but we did not write down a
definition of independence during the cycle. In practice, several
entries of one catalogue counted as one source, and separate projects of
the same publisher counted as separate sources. So SSDF and SP 800-53
counted as two NIST sources, CWE and CAPEC as two MITRE sources, and
ASVS, SAMM and DSOMM as three OWASP sources. Different teams write these
documents, but they share publishers and sometimes cite each other, so
when they agree, this is weaker evidence than agreement between
unrelated publishers. Until there is a written definition, the curator
judges independence case by case. A written definition, applied to all
decisions, will be part of the next version of the rule.

\subsubsection{5.3 The four decisions}\label{the-four-decisions}

Four candidate changes reached a decision (Table 3).

\needspace{8\baselineskip}

\textbf{Table 3.} The four change requests.

\begin{longtable}[]{@{}
  >{\raggedright\arraybackslash}p{(\linewidth - 6\tabcolsep) * \real{0.2500}}
  >{\raggedright\arraybackslash}p{(\linewidth - 6\tabcolsep) * \real{0.2500}}
  >{\raggedright\arraybackslash}p{(\linewidth - 6\tabcolsep) * \real{0.2500}}
  >{\raggedright\arraybackslash}p{(\linewidth - 6\tabcolsep) * \real{0.2500}}@{}}
\toprule\noalign{}
\begin{minipage}[b]{\linewidth}\raggedright
Change request
\end{minipage} & \begin{minipage}[b]{\linewidth}\raggedright
Content
\end{minipage} & \begin{minipage}[b]{\linewidth}\raggedright
Sources cited
\end{minipage} & \begin{minipage}[b]{\linewidth}\raggedright
Decision
\end{minipage} \\
\midrule\noalign{}
\endhead
\bottomrule\noalign{}
\endlastfoot
ACR-001 & Secure configuration baseline integrity & SSDF, ASVS, CIS
Controls, CWE & Tier B: three objectives with their practices,
mechanisms and four artefacts in the release and rollout domain \\
ACR-002 & Security requirements lifecycle management & SSDF, OWASP SAMM,
PCI Secure SLC, NIST SP 800-53, SAFECode & Tier B: one objective, two
practices and two mechanisms in the threat modelling and risk
disposition domain \\
ACR-003 & A batch of about 35 items that fitted weakly across domains &
No common concept & Not admitted: each item was already covered or was a
traceability problem \\
ACR-004 & Context-aware output encoding & ASVS, CWE, CAPEC, OWASP DSOMM,
MCP security documentation & Tier B: one objective, one practice and one
mechanism; the input validation domain widened to input and output \\
\end{longtable}

No change reached tier C. The three admitted changes added 21 elements
and the maintenance rule added four mechanisms, which took the model
from 234 to 259 elements and from 70 to 75 objectives. None of the
change requests came from a law or regulation. When we read the
regulations for their engineering controls, their duties landed on
objectives that the technical sources had already set up.

\subsubsection{5.4 One change, from sources to
elements}\label{one-change-from-sources-to-elements}

ACR-004 shows how a change moves from the sources into the model. During
the second round, the similarity substrate found a group of 69 items
from five sources whose closest elements were spread over an identity
practice, an input validation practice and an input validation
objective, and all of them were only weakly similar. The items included
ASVS requirements on output encoding for web pages (HTML), web addresses
(URLs) and JavaScript, CWE-79 on cross-site scripting, CAPEC patterns
for cross-site scripting, a DSOMM activity on context-aware output
encoding, and a theme in the MCP security documentation on injection
through rendered output in agent contexts.

The content was engineering practice, and it was different from the
existing input validation objective, which covers resistance to
injection on the input side. We rejected two other options. Renaming the
existing objective would have put output encoding items into an
objective about input. A new domain was not justified either, because
the items fitted the input validation domain once that domain also
covered output. So the change added ACO-IVF-008 (\emph{Context-Aware
Output Encoding and Rendering Safety}), ACP-IVF-007 and ACM-IVF-005, and
changed the domain contract. The graph passed validation again after the
change.

The decision was based on the group of 69 items found by the similarity
substrate, from five sources, and on the curator's reading of them.
After the change, 31 curated rows from seven sources (ASVS, CAPEC, CWE,
DSOMM, the OWASP Proactive Controls, the OWASP Top 10 and the SAFECode
guide on fundamental practices) were moved to ACO-IVF-008, 26 as primary
and 5 as secondary objective. In the MCP documentation the content
exists as a theme, with no curated item on rendering.

\begin{center}\rule{0.5\linewidth}{0.5pt}\end{center}

\subsection{6. From the model to what
applies}\label{from-the-model-to-what-applies}

AppSec Core holds content. It does not decide what applies to a given
change, and it does not deliver anything to an engineer or an agent.
Other components do those jobs. We describe them as they are built
today, because the way the jobs are split tells a reader what to expect
from the model.

\needspace{8\baselineskip}

\textbf{Table 4.} What each component does, as built today.

\begin{longtable}[]{@{}
  >{\raggedright\arraybackslash}p{(\linewidth - 6\tabcolsep) * \real{0.2500}}
  >{\raggedright\arraybackslash}p{(\linewidth - 6\tabcolsep) * \real{0.2500}}
  >{\raggedright\arraybackslash}p{(\linewidth - 6\tabcolsep) * \real{0.2500}}
  >{\raggedright\arraybackslash}p{(\linewidth - 6\tabcolsep) * \real{0.2500}}@{}}
\toprule\noalign{}
\begin{minipage}[b]{\linewidth}\raggedright
Component
\end{minipage} & \begin{minipage}[b]{\linewidth}\raggedright
Job
\end{minipage} & \begin{minipage}[b]{\linewidth}\raggedright
What it holds
\end{minipage} & \begin{minipage}[b]{\linewidth}\raggedright
What it does not do
\end{minipage} \\
\midrule\noalign{}
\endhead
\bottomrule\noalign{}
\endlastfoot
AppSec Core & Represents the engineering content & Objectives,
practices, mechanisms, artefacts, domains, links to source items & Holds
no risk levels, roles, phases or acceptance criteria, and selects
nothing \\
Manual & Puts the content into practice & Requirements with identifiers,
risk levels from L1 (lowest) to L3 (highest) and acceptance criteria;
user stories; assignments of requirements to roles and phases; expected
evidence; links from manual controls to objectives & Does not compute
what applies to a change; it tells a person how to classify the
application \\
Selection rules & Decide what applies from a declared context & Baseline
requirements per risk level, rules that switch on chapters from declared
concerns, technologies and changed files, and a filter by level & Do not
guess the risk level or interpret the text of the task \\
Delivery & Makes the result available & A Model Context Protocol server
with tools for selection, code-generation context, review scope and
verification & Does not write or check code, and does not check that
evidence exists \\
\end{longtable}

The manual is compiled, together with AppSec Core, into a knowledge
graph that the delivery server reads {[}15, 44{]}. In the current
release the manual has 284 requirements, 260 user stories, 1,521
assignments of requirements to roles and phases, and 99 metrics
{[}44{]}. A further layer of links, also written by the curator,
connects each manual requirement to one of 20 manual controls, and each
control to the AppSec Core objectives it matches, marked as exact,
partial or related.

\textbf{What selection needs.} The selection rules are built into the
delivery server {[}45{]} and follow a model declared in the manual's own
ontology. The requirements that apply are a baseline chapter for the
risk level, plus the chapters that the declared context switches on,
filtered by the levels at which each requirement applies. The inputs are
a risk level, which is required, and a declared context: concerns from a
closed list of 24, how exposed the application is, how sensitive its
data is, its technologies, and the files that a change touches. A person
decides the risk level with the classification rule in chapter 1 of the
manual, which adds up three scores for exposure, data and impact. The
server does not guess it, and when no context is declared it asks for
one instead of guessing from the text of the task. The tables that link
concerns, technologies and file patterns to chapters and domains are
written into the server and are part of its version.

\textbf{Getting a domain is not the same as deciding what applies.} The
selection tool returns identifiers of manual requirements and no AppSec
Core identifiers. A second tool, which prepares the context for code
generation, adds the whole AppSec Core domain that the declared concern
points to, through a fixed table. So the objectives that reach the user
are those of the domain, and not the ones reached from the selected
requirements through the links from requirements to controls to
objectives. Those links are used to trace the sources behind each
requirement. In short, what applies to a change is decided at the level
of manual requirements and chapters, and AppSec Core gives the domain
content and the trace to the sources.

\textbf{People and agents.} People and agents work from the same
requirements and the same identifiers, but they reach them in different
ways, and we do not assume that they decide in the same way. People read
the manual, organised by chapter and by role, with user stories,
checklists and the evidence expected at each phase. Agents call the
server, which returns identifiers with their descriptions, how to verify
them and the expected evidence, and an instruction to cite only
identifiers from the returned set. In the current release the two routes
are only loosely joined, because user stories link to practices and not
to requirement identifiers. The server also does not know what an agent
actually reads: in the study cited in Section 8.3, the delivered text
was so long that the agent often saved it to a file and read only part
of it {[}10{]}.

\begin{center}\rule{0.5\linewidth}{0.5pt}\end{center}

\subsection{7. Worked example: adding a third-party
dependency}\label{worked-example-adding-a-third-party-dependency}

We follow one change through the four components, using the published
delivery server, version 0.22.0, which serves manual release 1.16.0 and
knowledge graph 2.0.0 with 274 requirements {[}45{]}. We ran the calls
locally.

\textbf{Context.} A developer adds a third-party npm package to a web
service, which changes \texttt{package.json} and
\texttt{package-lock.json}. We declare risk level L2, which fits an
application with a medium total score for exposure, data and impact. We
chose this level for the example; it does not come from a
classification. We also declare the concern \texttt{supply\_chain}.

\textbf{Selection.} For this context the selection tool returned 32
requirements. Thirteen come from chapter 5 of the manual, on
dependencies, SBOM and software composition analysis, which the changed
file \texttt{package.json} switches on. They are in the dependency
category (DEP): DEP-001 to DEP-008 and DEP-010 to DEP-014. Ten come from
the continuous integration and delivery category (CIC) and nine from the
integrity category (INT), which the concern switches on. DEP-009 was
left out because it applies only at L3. Without the concern, the same
call returned only the thirteen DEP requirements, and with only a
description of the task and no declared context, it asked for more
input. The selected DEP requirements include an SBOM for every build
(DEP-001), composition analysis with a severity gate (DEP-002), pinned
versions (DEP-003), controlled registries (DEP-005) and formal approval
of new dependencies (DEP-006). They also include DEP-011 to DEP-014,
which are about dependencies on AI models and providers. These came in
for a plain library because selection works by chapter and not by how
relevant each item is. This is a limit of the current rules.

\textbf{From requirements to objectives.} All the DEP requirements of
the chapter are linked to one manual control, the inventory and analysis
of dependencies. That control matches three AppSec Core objectives:
ACO-SCBI-001 \emph{Dependency Inventory and SBOM Traceability} (exact),
ACO-SCBI-002 \emph{Dependency Risk Evaluation and Policy Gating} (exact)
and ACO-SCBI-003 \emph{Controlled Dependency and Image Sources}
(partial). Their practices are ACP-SCBI-001 \emph{Build-Linked SBOM
Generation}, ACP-SCBI-002 \emph{Automated Dependency and Image Risk
Gating} and ACP-SCBI-003 \emph{Approved Source and Registry Governance}.
Their mechanisms are ACM-SCBI-005 \emph{Build and Image Inventory
Generation}, ACM-SCBI-002 \emph{Automated Security Scanners} and
ACM-SCBI-006 \emph{Registry Allowlisting and Approved Source
Enforcement}. The context tool served the whole supply chain and build
integrity domain, which has seven objectives, so the user gets these
three objectives and four others from the same domain. Figure 2 shows
the whole path.

\begin{figure}
\centering
\includegraphics[width=1\linewidth,height=\textheight,keepaspectratio,alt={The worked example as one path: from a declared change to the requirements, the manual control, the three objectives with their practices, mechanisms and artefacts, and the source items behind each objective. One expected artefact is shown for each objective. The CIC and INT requirements, selected by the declared concern, are not followed further here.}]{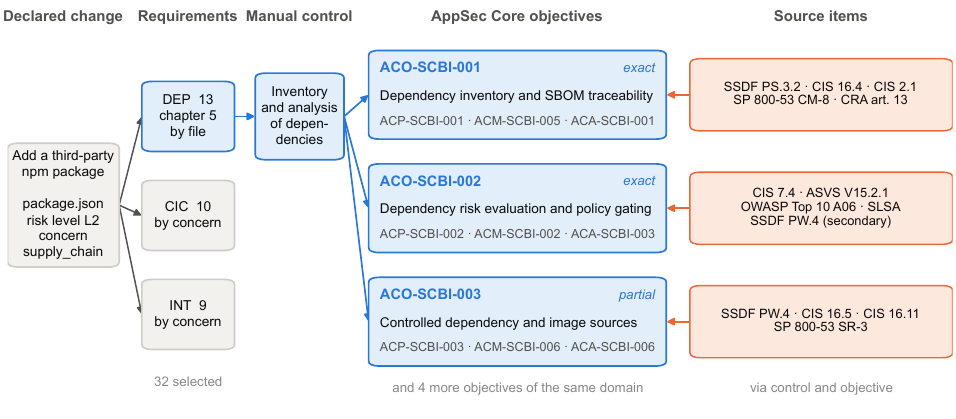}
\caption{The worked example as one path: from a declared change to the
requirements, the manual control, the three objectives with their
practices, mechanisms and artefacts, and the source items behind each
objective. One expected artefact is shown for each objective. The CIC
and INT requirements, selected by the declared concern, are not followed
further here.}
\end{figure}

\textbf{Expected evidence.} Each selected requirement has an evidence
pattern that names the expected types of artefact and repeats the
acceptance criterion. For this change the patterns point to the SBOM and
the list of dependencies, the composition analysis report, the approval
record and the release gate. In AppSec Core these match artefacts of the
supply-chain domain, for example ACA-SCBI-001 (the SBOM), ACA-SCBI-003
(the composition analysis report) and ACA-SCBI-006 (the approval
record).

\textbf{Sources behind the objectives.} The curated mappings show which
source items are behind the three objectives. SSDF practice PW.4, on
reusing secure software, and its tasks map to ACO-SCBI-003 with
ACO-SCBI-002 as secondary objective, and SSDF task PS.3.2, on collecting
provenance data, maps to ACO-SCBI-001. CIS safeguards 16.4 and 2.1 map
to ACO-SCBI-001, safeguards 16.5 and 16.11 to ACO-SCBI-003, and
safeguard 7.4 to ACO-SCBI-002. ASVS requirement V15.2.1, on keeping
components within their update and remediation time frames, maps to
ACO-SCBI-002 with bounded strength, and so does the SLSA guidance on
verifying dependencies. OWASP Top 10 category A06:2021, on vulnerable
and outdated components, maps to ACO-SCBI-002, and NIST SP 800-53
controls CM-8 and SR-3 map to ACO-SCBI-001 and ACO-SCBI-003. Among the
regulations, CRA article 13 maps to ACO-SCBI-001 with bounded strength.
In the curated mappings ACO-SCBI-002 is used by fifteen of the twenty
sources of group A, more than any other objective. None of the DEP
requirements cites a source directly, so every link from a requirement
to a source goes through the control and the objective.

\textbf{Use by an engineer and by an agent.} A developer who follows the
manual finds the same requirements in chapter 5, with user stories for
approving and checking a new dependency in the design phase, and the
expected artefacts. An agent that calls the server gets the requirement
identifiers, the domain content and an instruction to write a short
explanation that cites the identifiers it implemented and the evidence
it expects. The reviewer of the change can then check against the same
list. The server does not check that the evidence exists, and in the
current release two fields of the evidence patterns, the roles
responsible and the moment of verification, are empty for this chapter.

\textbf{What the example shows, and its limits.} For a declared context,
the components return a defined set of requirements, a domain of
objectives, practices, mechanisms and expected artefacts, and a trace to
the source items behind them, all with stable identifiers. The example
also shows three limits. What applies depends on a risk level and a
context that a person declares. The objectives served are those of a
whole domain, and not those reached from the selected requirements. And
selection by chapter can bring in requirements that do not fit this
particular change. A rule that would add related objectives from other
domains, such as the release objectives that a promoted artefact
reaches, has been proposed but is not built yet.

\begin{center}\rule{0.5\linewidth}{0.5pt}\end{center}

\subsection{8. Evidence}\label{evidence}

We report the evidence in three parts and keep them apart, because
evidence in one part does not carry over to another. Section 8.4 then
answers the research question.

\subsubsection{8.1 How well the model covers the
sources}\label{how-well-the-model-covers-the-sources}

\textbf{Stability of the structure.} Across the 31 sources, we did not
have to add any domain, element type or relation type. One domain
contract was widened from input to input and output (ACR-004), and the
model grew from 234 to 259 elements through additions inside the
existing structure. This holds for this set of sources and these
versions of them, inside the boundary, at the level of detail of the
model and under the change rule.

\textbf{How much of each source the model holds.} In the twenty sources
of group A, 2,458 of the 3,247 rows that are not headings or context
(75.7\%) landed on an element of the model, and the other 789 were
outside the boundary (Section 4.5). In the six sources of group B,
mapped at the level of selected articles and requirements, the share is
91.1\% (257 of 282). Table A1 gives the figures for each source.

\textbf{How many sources use each objective.} In the curated mappings of
group A, we count a source for an objective when one of its mapped rows
has that objective as primary or secondary anchor. The median objective
is used by six of the twenty sources, and the maximum is fifteen.
Sixty-one of the 75 objectives are used by three or more sources, six by
exactly one source, and two by none. If we count only strong and direct
rows, the median drops to five, 55 objectives are used by three or more
sources, seven by one and four by none. If we add the six sources of
group B, the median rises to seven. Table 5 gives the distribution, and
Figure 3 shows it by domain.

\textbf{Table 5.} Number of group A sources that use each objective
(primary or secondary anchor, curated mappings).

\begin{longtable}[]{@{}
  >{\raggedright\arraybackslash}p{(\linewidth - 30\tabcolsep) * \real{0.2105}}
  >{\raggedright\arraybackslash}p{(\linewidth - 30\tabcolsep) * \real{0.0526}}
  >{\raggedright\arraybackslash}p{(\linewidth - 30\tabcolsep) * \real{0.0526}}
  >{\raggedright\arraybackslash}p{(\linewidth - 30\tabcolsep) * \real{0.0526}}
  >{\raggedright\arraybackslash}p{(\linewidth - 30\tabcolsep) * \real{0.0526}}
  >{\raggedright\arraybackslash}p{(\linewidth - 30\tabcolsep) * \real{0.0526}}
  >{\raggedright\arraybackslash}p{(\linewidth - 30\tabcolsep) * \real{0.0526}}
  >{\raggedright\arraybackslash}p{(\linewidth - 30\tabcolsep) * \real{0.0526}}
  >{\raggedright\arraybackslash}p{(\linewidth - 30\tabcolsep) * \real{0.0526}}
  >{\raggedright\arraybackslash}p{(\linewidth - 30\tabcolsep) * \real{0.0526}}
  >{\raggedright\arraybackslash}p{(\linewidth - 30\tabcolsep) * \real{0.0526}}
  >{\raggedright\arraybackslash}p{(\linewidth - 30\tabcolsep) * \real{0.0526}}
  >{\raggedright\arraybackslash}p{(\linewidth - 30\tabcolsep) * \real{0.0526}}
  >{\raggedright\arraybackslash}p{(\linewidth - 30\tabcolsep) * \real{0.0526}}
  >{\raggedright\arraybackslash}p{(\linewidth - 30\tabcolsep) * \real{0.0526}}
  >{\raggedright\arraybackslash}p{(\linewidth - 30\tabcolsep) * \real{0.0526}}@{}}
\toprule\noalign{}
\begin{minipage}[b]{\linewidth}\raggedright
Sources
\end{minipage} & \begin{minipage}[b]{\linewidth}\raggedright
0
\end{minipage} & \begin{minipage}[b]{\linewidth}\raggedright
1
\end{minipage} & \begin{minipage}[b]{\linewidth}\raggedright
2
\end{minipage} & \begin{minipage}[b]{\linewidth}\raggedright
3
\end{minipage} & \begin{minipage}[b]{\linewidth}\raggedright
4
\end{minipage} & \begin{minipage}[b]{\linewidth}\raggedright
5
\end{minipage} & \begin{minipage}[b]{\linewidth}\raggedright
6
\end{minipage} & \begin{minipage}[b]{\linewidth}\raggedright
7
\end{minipage} & \begin{minipage}[b]{\linewidth}\raggedright
8
\end{minipage} & \begin{minipage}[b]{\linewidth}\raggedright
9
\end{minipage} & \begin{minipage}[b]{\linewidth}\raggedright
10
\end{minipage} & \begin{minipage}[b]{\linewidth}\raggedright
11
\end{minipage} & \begin{minipage}[b]{\linewidth}\raggedright
12
\end{minipage} & \begin{minipage}[b]{\linewidth}\raggedright
13
\end{minipage} & \begin{minipage}[b]{\linewidth}\raggedright
15
\end{minipage} \\
\midrule\noalign{}
\endhead
\bottomrule\noalign{}
\endlastfoot
Objectives & 2 & 6 & 6 & 5 & 8 & 8 & 7 & 10 & 6 & 6 & 4 & 3 & 2 & 1 &
1 \\
\end{longtable}

\begin{figure}
\centering
\includegraphics[width=1\linewidth,height=\textheight,keepaspectratio,alt={Number of group A sources that use each of the 75 objectives, one dot per objective, by domain. The dashed line is the median. The shaded area marks the objectives used by two sources or fewer.}]{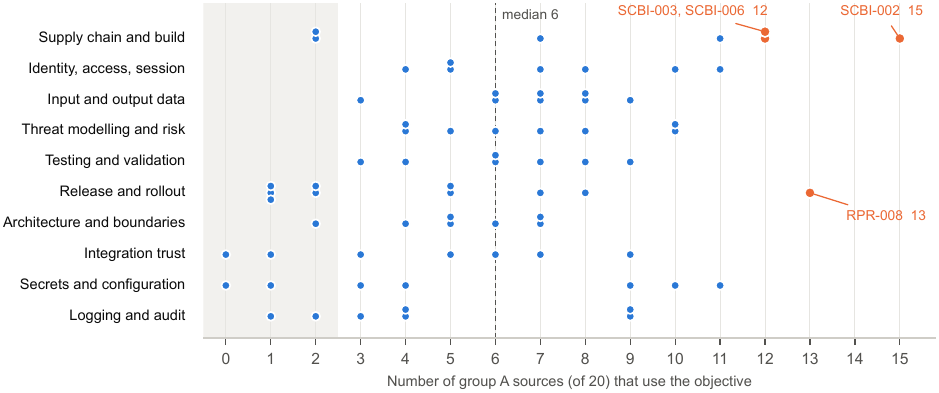}
\caption{Number of group A sources that use each of the 75 objectives,
one dot per objective, by domain. The dashed line is the median. The
shaded area marks the objectives used by two sources or fewer.}
\end{figure}

The objectives used by most sources are in the supply-chain and
configuration areas: ACO-SCBI-002 (fifteen sources), ACO-RPR-008
\emph{Secure Configuration Baseline Integrity} (thirteen), and
ACO-SCBI-003 and ACO-SCBI-006 \emph{Artifact Attestation and Provenance}
(twelve each). A high count means that the objective is linked to many
sources at once. This supports the idea in Section 1.2: different
publications keep asking for the same things. The objectives with one
source or none rest on the practice written in the manual and not on the
outside sources. The two that no curated row uses are ACO-ITS-007
\emph{Integration Trust and Service Interaction Assurance} and
ACO-SPC-006 \emph{Secret Change Auditability and Governance}. The six
with one source are ACO-ITS-006, ACO-RPR-005, ACO-RPR-006, ACO-RPR-007,
ACO-SLG-006 and ACO-SPC-004. We list them so that they can be reviewed:
when more sources are added, they will either be confirmed or removed.

\textbf{Traceability of elements.} If we count only the curated anchor
fields of the 26 curated sources, 194 of the 202 objectives, practices
and mechanisms are linked to at least one source row: 73 of 75
objectives, 68 of 69 practices and 53 of 58 mechanisms. With the twenty
sources of group A alone, the figure is 187. The eight elements without
a link are the two objectives named above, the four mechanisms added
under the maintenance rule (ACM-RPR-005, ACM-SCBI-006, ACM-SLG-005 and
ACM-SLG-006), and a practice and a mechanism added by ACR-001
(ACP-RPR-010 and ACM-RPR-010). For these elements, the trace exists only
in the change records or in the manual. All 57 artefacts are expected by
at least one objective that has a source row, so their trace to the
sources is indirect but complete. An earlier count of 192 linked
elements {[}13{]} refers to mapping state 1, before the three elements
of ACR-004 were linked, and it also counted similarity candidates as
links.

\textbf{Sources on AI systems and agents.} Ten of the 31 sources are
about AI systems or AI agents, and we treated them at two depths. The
five in group A were mapped row by row: the security documentation of
the Model Context Protocol, three OWASP documents on MCP servers and
their risks, and the ENISA framework of good cybersecurity practices for
AI. Their 51 curated rows include 40 mapped rows. The MCP documents
describe risks such as token passthrough and audience binding, confused
deputies, scope creep, tool poisoning, prompt injection through tool
output, memory poisoning and sharing too much context. These landed on
existing objectives, mostly in the identity, access and session trust
domain (15 of the 40 rows), the input and output data safety domain (7)
and the supply chain and build integrity domain (6), where tool
poisoning is treated as a dependency risk. The ENISA framework was
mapped only at the level of its three layers, with bounded strength. The
MCP security documentation is one of the sources cited for ACR-004,
although only as a theme (Section 5.4). The five sources of group C,
which we added last as a deliberate test, were first compared with the
model through the similarity substrate. Of their 404 items, 267 (66.1\%)
had at least one link above the threshold. The other 137 went through an
automated review with several language models, and the curator then
reviewed, completed and corrected the result. The review found that
existing elements could hold these items without a new practice
{[}14{]}. The criterion set before the round asked for 70\% of the items
to have a link, so for group C the result rests on that review and not
on the links. For the material on agent security in group A, the curated
mappings show that the existing structure holds it.

\subsubsection{8.2 Are the mappings
correct?}\label{are-the-mappings-correct}

One curator produced and reviewed the mappings, and the same person
wrote the manual that the model was taken from. The model and the manual
are open, but there has not yet been an independent expert review,
because there has been no funding or formal setting for it, and we have
not measured agreement between raters. We report two kinds of check: one
on the curated mappings, and one on the similarity substrate, which
corrects figures reported earlier.

The first is a three-way cross-reference check on the curated mappings.
For each of its tasks, SSDF 1.1 publishes references to other sources,
among them NIST SP 800-53, OWASP SAMM, the PCI Secure SLC standard and
SAFECode guides. Because we mapped these sources separately from SSDF, a
reference that SSDF publishes can serve as an outside check: if SSDF
says that one of its tasks matches an item of SAMM, both items should
land on the same objective. We ran this check again on mapping state 2,
over 138 pairs of a task and a referenced item at the same level of
detail, and compared the result with a baseline where the referenced
items were shuffled within each source. The two items had the same
primary objective in 42.8\% of pairs, against 8.7\% for shuffled pairs,
and shared at least one primary or secondary objective in 70.3\%,
against 21.2\%. Grouped by task and referenced source, which was the
unit of an earlier run of the check, 81.0\% of the 84 groups shared an
objective, against 29.4\% for shuffled pairs, and 88.1\% shared an
objective or a domain, against 46.6\%. In 1,000 shuffles, no baseline
reached the observed values (Figure 4, left). So the check shows
agreement well above chance, with two limits. It covers only the sources
that SSDF references, and the earlier run of the check was also used to
correct some mappings, so it is not independent of them.

The second kind of check compared the similarity substrate with the same
SSDF references and with the cross-references of the Secure Controls
Framework (SCF) {[}46{]}. Earlier preprints reported agreement of 100\%
for 35 SSDF items and 97.46\% for 355 source items in the SCF comparison
{[}13, 14{]}. Those figures count an item as agreeing when any of its
links shares any domain with any link of an item that the outside
reference pairs with it, and the links of each item cover about half of
the ten domains. When the pairs are shuffled within each pool, the same
measure gives 98.2\% and 97.3\%. So these figures do not separate the
substrate from chance (Figure 4, right), and we do not use them as
validation. A stricter measure, which asks for the same top-similarity
objective on both sides, gives 19.7\% of 142 SSDF pairs and 10.0\% of
5,652 SCF pairs, which shows how loose the substrate links are (Section
4.2). The difference between the two layers is useful in itself: the
curated mappings agree with an outside reference far better than the
similarity links, and none of the coverage, support or traceability
counts uses the similarity links.

\begin{figure}
\centering
\includegraphics[width=1\linewidth,height=\textheight,keepaspectratio,alt={Left: the three-way check on the curated mappings, observed against 1,000 shuffles of the referenced items (gray band: 95\% of the shuffles; gray dot: their mean). Right: the shared-domain measure on the similarity links, per item, as reported in earlier preprints, against 200 shuffles of the pairs.}]{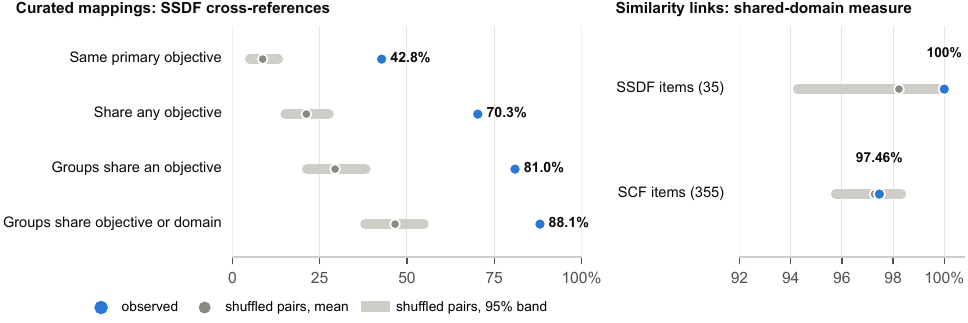}
\caption{Left: the three-way check on the curated mappings, observed
against 1,000 shuffles of the referenced items (gray band: 95\% of the
shuffles; gray dot: their mean). Right: the shared-domain measure on the
similarity links, per item, as reported in earlier preprints, against
200 shuffles of the pairs.}
\end{figure}

So today, whether each mapping is right rests on the recorded reason for
each row, on the review by one curator, on the three-way check, and on
the internal spot check in Section 4.2. An independent review of a
sample of curated rows, with a measure of agreement, is the most
important next step here. Whether each mapping is right is also a
different question from whether the model does its job. Section 8.3
reports the evidence on that second question.

\subsubsection{8.3 Use in practice}\label{use-in-practice}

Section 7 shows that the components can turn a declared context into a
defined set of requirements, objectives, practices, mechanisms and
expected evidence, with stable identifiers and a trace to the sources.

\textbf{Fit for purpose.} The purpose of the model is to tell an
engineer or an agent what security a change needs. We have two kinds of
evidence on this, and they are of a different kind. The first is a
demonstration of use that was reviewed by others. The second is a
measurement of results in code generation, where outside benchmarks
exist.

\textbf{A reviewed demonstration of use.} In a separate paper, accepted
at a peer-reviewed workshop on modelling for cybersecurity {[}16{]}, we
took a change to authentication middleware and asked the published graph
what the change must meet. The change belongs to the identity, access
and session trust domain, and the query returned the seven objectives of
that domain. When we opened one of them, access abuse detection and
auditability (ACO-IAT-006, Box 1), we got one practice, two mechanisms
and three expected artefacts: a log stream, a policy report and an alert
policy. The answer is a closed set: in the graph every relation stays
inside its domain, so a query on a domain cannot go past it. The answer
also shows shared evidence: the log stream (ACA-IAT-003) is expected by
six of the seven objectives, so one log stream can serve as evidence for
all six. We checked these facts again on the current graph, which is the
same file. That paper also describes what a closed set costs: a duty
that comes from reading several provisions together, which is common in
NIS2, reaches the engineer one claim at a time, and a change that spans
domains needs one query per domain.

The same graph also answers the opposite question, from a source to the
work it asks for. The workshop paper showed this for a regulation, but
through the similarity layer, so here we checked it on the curated rows
instead. The ten risks of the OWASP MCP Top 10 land, as primary or
secondary objective, on eleven objectives in seven domains: identity,
access and session trust; input and output data safety; release and
rollout; supply chain and build integrity; security logging and audit;
secrets and protected configuration; and threat modelling and risk
disposition. So a request such as ``make this MCP server address the MCP
Top 10'' becomes a closed set of objectives, each with its practices,
mechanisms and expected artefacts. In both directions this is a
demonstration of use, not a measurement.

\textbf{Measured results in code generation.} We measured results in
code generation, where public benchmarks run security tests on generated
code. In a separate study we gave requirements derived from the manual
and AppSec Core to the code generator, and measured the result on two
benchmarks, DualGauge {[}47{]} and BaxBench {[}48{]} {[}10{]}. On
DualGauge, where the generator was a coding agent, the share of security
tests passed went from 77.4\% to 93.0\% (287 and 345 of 371 tests), and
the share of tasks that passed all of them went from 44.1\% to 78.0\%.
The authors of BaxBench also define what they call an ``unrealistic
oracle security reminder'': a prompt where the developer knows in
advance every weakness of the scenario and asks the model to avoid them
{[}48{]}. In the BaxBench direct-call experiment, an agent selected the
requirements and the benchmark generated the code with one call to the
model. There, the share of functionally correct solutions with no
successful exploit was 65\% with the benchmark's baseline prompt, 86\%
with the delivered requirements and 85\% with its oracle prompt. So the
delivered requirements reached a similar level without knowing which
weaknesses would be tested.

This evidence is narrow in two ways. First, it covers only code
generation, and not the other work that the model is meant to support,
such as design, review, testing or release. Second, even in code
generation, the benchmarks only measure whether a small set of
weaknesses is avoided. Each BaxBench scenario runs one to four security
tests, and the whole benchmark checks thirteen types of weakness. Much
of what the model brings in is outside what the tests can see. In the
generated code, controls such as rate limiting, generic error handling,
disabled documentation of the application programming interface (API),
limits on field length and clean logs appeared mostly or only when the
requirements were delivered, and no test scores them. The study counted
whether they were present by text search, not whether they work
{[}10{]}. The benchmarks can even penalise these controls: in seven of
the 23 DualGauge tasks that became secure, a functional test failed
because the code enforced a requirement that the test did not expect.

\textbf{The study.} Its design, its pre-specified outcome and its limits
are reported in {[}10{]}. It used an earlier version of the delivery
server (0.10.1, with knowledge graph 1.5.0), which took the scope from
the text of the task and not from a declared context.

The two kinds of evidence answer different questions. The demonstration
shows that the model gives a closed, typed answer at a point of change.
The study shows that delivering requirements derived from the manual and
AppSec Core changes the code that is generated.

\subsubsection{8.4 Answer to the research
question}\label{answer-to-the-research-question}

Inside its boundary, a bounded model of engineering work can hold most
of the content of many different sources: three quarters of the rows of
the technical sources that are not headings or context landed on the
model, and we did not need a new domain, element type or relation. It
keeps their traceability: 194 of its 202 objectives, practices and
mechanisms are linked to source items, and every artefact is reached
through an objective. And it helps decide what applies to a change: for
the domain of a change, and for a source, it gives a closed, typed set
of elements, while the choice of requirements for a given change is made
by the manual and the selection rules (Section 6).

\begin{center}\rule{0.5\linewidth}{0.5pt}\end{center}

\subsection{9. Related work}\label{related-work}

Several kinds of work connect security guidance across sources. We
describe each one briefly and say how AppSec Core relates to it, looking
only at what we could check in the published material.

\textbf{OWASP OpenCRE} {[}49{]} is the closest existing work, and it
starts from the same basic idea as AppSec Core: the many publications on
application security keep coming back to a limited set of things, in
different words. OpenCRE captures this as Common Requirement
Enumerations (CREs), each one ``a normalized security requirement'' to
which standards link their sections, so that practitioners can see how
topics line up across sources. It has stable identifiers, a hierarchy of
topics, an API, an overlap analysis between two standards and its own
MCP server, and it is well established and maintained by a community.
Its hierarchy covers much the same ground as AppSec Core. Under
development processes it has design, build, supply chain and
verification, and under technical controls it has authentication, input
and output protection and logging.

The two differ mainly in purpose. OpenCRE is built to connect standards
to each other through common requirements, so its links are the core of
what it offers. It organises its ground by subject: a CRE says what a
requirement is about, and conditions, practices and mechanisms can sit
side by side under the same topic. AppSec Core is built to describe the
engineering work itself: what must be true, which practice gets there,
which mechanism puts it in place and which evidence shows it. Its
mappings come second. They show how much of each source the model covers
and whether the sources support it, and they are not meant as a
crosswalk. Neither model holds the lifecycle view, with its phases and
roles; in our case the manual adds it.

This difference in purpose is why we did not just reuse OpenCRE as the
reference for the manual. To check how much a manual covers, we needed a
fixed and bounded set of elements, so that coverage is a stable number.
We needed elements typed by role, so that a gap can be stated precisely.
And we needed the sources that the manual has to answer to, with a
recorded strength and recorded exclusions for each link. In the version
of OpenCRE we looked at (repository commit \texttt{ace5f0b} and live
catalogue, 6 November 2026), CREs are not typed by role, links record no
strength (its map analysis ranks the paths that link two standards by a
computed score), and 14 of our 31 sources are among its linked
standards, in some cases in other versions. OpenCRE is designed for a
broader purpose and serves it well. In future work its links are a
natural extra source for AppSec Core, and AppSec Core objectives could
in turn be linked to CREs.

\textbf{Crosswalks and the NIST National Online Informative References
(OLIR) programme} {[}50, 51{]} define how to map a reference document to
a focal document, with relationship types from set theory (subset of,
intersects with, equal, superset of, not related to) or supportive
relationships, a reason and an optional strength. AppSec Core can act as
a common focal document for all sources, and its strength labels (strong
or direct, and bounded) are a coarser scale than the OLIR strength. The
similarity links of the design cycle have also been published in the
OLIR format {[}14{]}.

\textbf{OSCAL}, the Open Security Controls Assessment Language of NIST
{[}52{]}, represents control catalogues in a machine-readable form, and
profiles that select and tailor controls from a catalogue for a given
system. A profile is the closest standard counterpart of the selection
described in Section 6. AppSec Core and the selection rules could be
expressed in these terms, with the model as a catalogue and each
declared context as a profile.

\textbf{Control frameworks} such as the Secure Controls Framework
{[}46{]}, which maps more than 200 laws and frameworks to more than
1,500 controls with the NIST set-theory relationships, and the Cloud
Security Alliance Cloud Controls Matrix {[}53{]} are common lists of
controls for governance and compliance. Their controls speak to the
organisation as a whole, at a level above a single code change, while
AppSec Core speaks to the engineering work of a team.

\textbf{ISO/IEC 27034} {[}54{]} describes how an organisation keeps a
framework of application security controls and selects, for each
application, the part of it that applies. This selection is close in
spirit to the selection described in Section 6. The standard defines a
method for one organisation, not a shared set of elements.

\textbf{Security ontologies} such as the Unified Cybersecurity Ontology
(UCO) {[}55{]}, STUCCO {[}56{]} and MITRE D3FEND {[}57{]} model attacks,
weaknesses, indicators and countermeasures for detection, analysis and
defence. They are about operational security knowledge, not about the
duties of development work, and they can be joined with AppSec Core
through the CWE and CAPEC identifiers that several of them use. In
requirements engineering, reusable security knowledge has been organised
as ontologies, patterns and catalogues {[}58{]}, including security
ontologies built from control catalogues such as IT-Grundschutz and
mapped to ISO/IEC 27001 {[}59{]} and reusable templates for security
requirements {[}60{]}. AppSec Core shares the aim of reusable security
knowledge, and it adds links from its elements to the items of current
standards and regulations.

We built the model step by step, in the style of ontology engineering
methods such as NeOn {[}61{]} and LOT {[}62{]}, and the check against
the sources is reported as a design science cycle {[}63{]} in {[}14{]}.
We do not present the change rule as a general method of ontology
engineering. It is the rule this model needed in order to change under
outside evidence and still stay small.

\begin{center}\rule{0.5\linewidth}{0.5pt}\end{center}

\subsection{10. Limitations}\label{limitations}

\textbf{One curator and no independent review.} One person produced the
model, the mappings and the change decisions, and the same person wrote
the manual that the model was taken from. The model and the manual are
open for anyone to review, but no funded or formal review has taken
place. A language model suggested some corrections to the mappings, and
an AI agent made some later edits, all reviewed by the author. No
independent expert has reviewed the mappings or the decisions.

\textbf{Uneven depth of mapping.} The twenty sources of group A were
mapped by rules and review, with about half of the rows inherited from a
parent or a section. The six sources of group B were mapped with a
heuristic at the level of selected articles and requirements, and the
five sources of group C have similarity links and a review of the items
left without a link.

\textbf{Coverage and correctness.} The check shows that the content of
the sources could be placed in the model, and that the model needed no
new structure to hold it. It does not show that each mapping is right,
or that an application that meets every objective that applies is
secure.

\textbf{Reproducibility and repeatability.} The similarity substrate is
computed in a fixed way and can be computed again from the published
artefacts. The curated mappings record the decisions of one curator, and
another curator might decide differently.

\textbf{Incomplete records.} The curated files do not record, row by
row, whether excluded content went to the manual or out of application
security, and the summary counts stored in thirteen of the files no
longer match their rows. The counts in this paper are computed from the
rows.

\textbf{Independence of sources.} The change rule has no written
definition of independence, so the curator judges it case by case. A
written definition is planned for the next version of the rule.

\textbf{Known mapping problems.} The audit behind mapping state 2 found
mappings inherited from a parent without a review of each item, and
provisional anchors that nobody looked at again after ACR-001. These
were fixed only where they were found, and the heuristic mapping of
group B was not revised.

\textbf{Sources and versions.} The results hold for these 31 sources in
the versions listed. Several sources have newer versions, and further
sources, in particular on AI and machine-learning systems, may need new
structure.

\textbf{No study with engineers, and no comparison in use.} The paper
argues that one reference cuts the work of finding what applies to a
change, but it reports no measurement of that effort with engineers. We
have also not compared, in real use, working from AppSec Core with
working from broader catalogues such as OpenCRE.

\begin{center}\rule{0.5\linewidth}{0.5pt}\end{center}

\subsection{11. Conclusion}\label{conclusion}

AppSec Core writes down once, in engineering terms and inside a declared
boundary, what application security asks for during development: the
conditions that must be true, the ways to meet them and the evidence
that they were met. The curated mappings link 194 of its 202 objectives,
practices and mechanisms to source items, and these links show how far
the sources cover and support the model. We built the model from a
practitioner manual and checked it against 31 outside sources. It kept
its ten domains, changed three times under a set rule and widened one
domain, and in the curated mappings most of its objectives are used by
several sources. Together with a manual, selection rules and a delivery
server, which are separate components, it turns a declared change into a
set of requirements, objectives, practices, mechanisms and expected
evidence that engineers and agents can share. The model covers the
sources broadly and with traceability, and it gives a closed answer at a
point of change, for the domain of a change or for a source. Delivering
requirements derived from the manual and the model improved the security
of generated code on two benchmarks. The next steps are an independent
review of a sample of mappings, a written definition of source
independence, selection rules that reach objectives from the selected
requirements, and a study of the effort of engineers who work from one
reference.

\begin{center}\rule{0.5\linewidth}{0.5pt}\end{center}

\subsection{Availability}\label{availability}

The ontology (OWL 2 export, SHACL constraints and domain contracts) and
the artefacts of the design cycle (decomposed sources, similarity
substrate, OLIR crosswalks of the similarity links, and cross-reference
audits) are in release v2.0.0 of the public repository (the release
number of the repository is independent of the version of the model),
archived on figshare under CC BY 4.0 {[}17{]}. The change records of the
four decisions are part of that archived release. The compiled runtime
of the manual is deposited on Zenodo {[}44{]}, and the delivery server
is published on npm {[}45{]}.

\subsection{AI use statement}\label{ai-use-statement}

Generative language models helped draft and edit the text of this paper
and helped check the data behind it. While the mappings were built, a
language model acting as a reviewer suggested some of the 32 corrections
of April 2026, and an AI agent carried out the targeted re-anchoring of
October 2026. The items of group C left without a similarity link went
through an automated review with several language models. In all cases
the author reviewed, completed and approved the result. The similarity
substrate uses a sentence-embedding model and no generative model. The
author is responsible for all claims, decisions and attributions.

\begin{center}\rule{0.5\linewidth}{0.5pt}\end{center}

\subsection{References}\label{references}

{[}1{]} M. Souppaya, K. Scarfone, D. Dodson. \emph{Secure Software
Development Framework (SSDF) Version 1.1: Recommendations for Mitigating
the Risk of Software Vulnerabilities}. NIST SP 800-218, February 2022.

{[}2{]} OWASP. \emph{Application Security Verification Standard 5.0.0},
May 2025.

{[}3{]} OpenSSF. \emph{Supply-chain Levels for Software Artifacts (SLSA)
v1.0}, April 2023.

{[}4{]} Center for Internet Security. \emph{CIS Critical Security
Controls Version 8.1.2}, March 2025.

{[}5{]} MITRE. \emph{Common Weakness Enumeration (CWE) 4.19.1}, Software
Development view (CWE-699), January 2026.

{[}6{]} MITRE. \emph{Common Attack Pattern Enumeration and
Classification (CAPEC) 3.9}, January 2023.

{[}7{]} Regulation (EU) 2024/2847 of the European Parliament and of the
Council of 23 October 2024 on horizontal cybersecurity requirements for
products with digital elements (Cyber Resilience Act). OJ L, 2024/2847,
20.11.2024.

{[}8{]} Directive (EU) 2022/2555 of the European Parliament and of the
Council of 14 December 2022 on measures for a high common level of
cybersecurity across the Union (NIS 2 Directive). OJ L 333, 27.12.2022,
pp.~80--152.

{[}9{]} Regulation (EU) 2022/2554 of the European Parliament and of the
Council of 14 December 2022 on digital operational resilience for the
financial sector. OJ L 333, 27.12.2022, pp.~1--79. Known as the Digital
Operational Resilience Act (DORA).

{[}10{]} P. Farinha. \emph{Applying Security by Design at the Point of
Execution: How Governed Security Requirements Affect the Security of
AI-Generated Code}. arXiv:2610.10659, 2026.

{[}11{]} P. Farinha. \emph{SbD-ToE Manual: Security by Design across the
software lifecycle}, release v1.17.1.
\url{https://github.com/SbD-ToE/sbd-toe-manual} and
\url{https://www.securitybydesign.dev}, CC BY-SA 4.0.

{[}12{]} P. Farinha. \emph{AppSec Core: A Normalized Ontology for
Security Requirements Across Heterogeneous Frameworks}. OSF preprint,
2026. \url{doi:10.17605/OSF.IO/WG8PV}.

{[}13{]} P. Farinha. \emph{AppSec Core v1: A Formalized Normalization
Ontology for Application Security}. OSF preprint, 2026.
\url{doi:10.17605/OSF.IO/U9CRD}.

{[}14{]} P. Farinha. \emph{Pressure-Testing AppSec Core: A Design
Science Cycle for Bounded-Ontology Evolution Under Heterogeneous
Application-Security Sources}. OSF preprint, 2026.
\url{doi:10.17605/OSF.IO/3E8G5}.

{[}15{]} P. Farinha. \emph{Coverage-Preserving Compilation v2: A
31-Source Pipeline with Joint Manual and Knowledge-Graph Production}.
OSF preprint, 2026. \url{doi:10.17605/OSF.IO/TXW8P}.

{[}16{]} P. Farinha. \emph{What Must Be Done Here? Using a Bounded
Application-Security Ontology at the Point of Change}. Semantic Shields
3: International Workshop on Modeling for Cybersecurity, JOWO 2026, to
appear in CEUR-WS. arXiv:2611.01127, 2026.

{[}17{]} P. Farinha. \emph{AppSec Core Ontology Research v2.0.0: Curated
Research Program Release}. figshare, 2026.
\url{doi:10.6084/m9.figshare.32307669}.

{[}18{]} W3C OWL Working Group. \emph{OWL 2 Web Ontology Language
Document Overview (Second Edition)}. W3C Recommendation, 2012.

{[}19{]} H. Knublauch, D. Kontokostas (eds.). \emph{Shapes Constraint
Language (SHACL)}. W3C Recommendation, 2017.

{[}20{]} M. Poveda-Villalón, A. Gómez-Pérez, M. C. Suárez-Figueroa.
``OOPS! (OntOlogy Pitfall Scanner!): An On-line Tool for Ontology
Evaluation''. \emph{International Journal on Semantic Web and
Information Systems} 10(2):7--34, 2014.
\url{doi:10.4018/ijswis.2014040102}.

{[}21{]} Joint Task Force. \emph{Security and Privacy Controls for
Information Systems and Organizations}. NIST SP 800-53 Rev.~5, September
2020; release 5.2.0 of the OSCAL catalogue, August 2025.

{[}22{]} OWASP. \emph{Software Assurance Maturity Model (SAMM) v2.1},
2024.

{[}23{]} OWASP. \emph{DevSecOps Maturity Model (DSOMM)}, model version
4.2.0. \url{https://dsomm.owasp.org}, retrieved 5 April 2026.

{[}24{]} OWASP. \emph{OWASP Top 10:2021}, 2021.

{[}25{]} OWASP. \emph{OWASP Top 10 Proactive Controls 2018}, 2018.

{[}26{]} PCI Security Standards Council. \emph{Software Security
Framework: Secure Software Lifecycle (Secure SLC) Requirements and
Assessment Procedures, Version 1.1}, February 2021.

{[}27{]} SAFECode. \emph{Practical Security Stories and Security Tasks
for Agile Development Environments}, July 2012.

{[}28{]} SAFECode. \emph{Fundamental Practices for Secure Software
Development: Essential Elements of a Secure Development Lifecycle
Program}, Third Edition, March 2018.

{[}29{]} SAFECode. \emph{Software Integrity Controls: An Assurance-Based
Approach to Minimizing Risks in the Software Supply Chain}, June 2010.

{[}30{]} Model Context Protocol. \emph{Authorization} (specification
revision 2025-11-25) and \emph{Security Best Practices} (specification
revision 2025-06-18).
\url{https://modelcontextprotocol.io/specification}, retrieved 4 April
2026.

{[}31{]} OWASP. \emph{OWASP MCP Top 10} (beta), 2025.
\url{https://owasp.org/www-project-mcp-top-10/}.

{[}32{]} OWASP GenAI Security Project. \emph{A Practical Guide for
Secure MCP Server Development}, version 1.0, February 2026.
\url{https://genai.owasp.org/resource/a-practical-guide-for-secure-mcp-server-development/}.

{[}33{]} OWASP GenAI Security Project. \emph{CheatSheet -- A Practical
Guide for Securely Using Third-Party MCP Servers 1.0}, 2025.

{[}34{]} ENISA. \emph{Multilayer Framework for Good Cybersecurity
Practices for AI}, June 2023.

{[}35{]} Regulation (EU) 2016/679 of the European Parliament and of the
Council of 27 April 2016 on the protection of natural persons with
regard to the processing of personal data and on the free movement of
such data, and repealing Directive 95/46/EC (General Data Protection
Regulation). OJ L 119, 4.5.2016, pp.~1--88.

{[}36{]} PCI Security Standards Council. \emph{Payment Card Industry
Data Security Standard: Requirements and Testing Procedures, v4.0.1},
June 2024.

{[}37{]} U.S. Department of Health and Human Services. \emph{HIPAA
Security Rule}, 45 CFR Part 160 and Part 164, Subparts A and C.

{[}38{]} MITRE. \emph{ATLAS: Adversarial Threat Landscape for
Artificial-Intelligence Systems}, data release v5.6.0.
\url{https://atlas.mitre.org} and
\url{https://github.com/mitre-atlas/atlas-data}, retrieved 7 May 2026.

{[}39{]} OWASP. \emph{OWASP Top 10 for LLM Applications 2025}, November
2024.

{[}40{]} OWASP. \emph{OWASP Machine Learning Security Top Ten}, 2023
edition, draft v0.3.
\url{https://owasp.org/www-project-machine-learning-security-top-10/}.

{[}41{]} A. Vassilev, A. Oprea, A. Fordyce, H. Anderson, X. Davies, M.
Hamin. \emph{Adversarial Machine Learning: A Taxonomy and Terminology of
Attacks and Mitigations}. NIST AI 100-2 E2025, March 2025.
\url{doi:10.6028/NIST.AI.100-2e2025}.

{[}42{]} NIST. \emph{Artificial Intelligence Risk Management Framework
(AI RMF 1.0)}. NIST AI 100-1, January 2023.
\url{doi:10.6028/NIST.AI.100-1}.

{[}43{]} N. Reimers, I. Gurevych. ``Sentence-BERT: Sentence Embeddings
using Siamese BERT-Networks''. In \emph{Proc. EMNLP-IJCNLP 2019},
pp.~3982--3992. \url{doi:10.18653/v1/D19-1410}.

{[}44{]} P. Farinha. \emph{SbD-ToE Manual: compiled knowledge-graph
runtime 2.1.0 (manual v1.17.1)}. Zenodo, 2026.
\url{doi:10.5281/zenodo.23614726}.

{[}45{]} P. Farinha. \emph{SbD-ToE MCP server}, npm package
\texttt{@shiftleftpt/sbd-toe-mcp}, version 0.22.0, 2026.
\url{https://github.com/SbD-ToE/sbd-toe-mcp}.

{[}46{]} Secure Controls Framework Council. \emph{Secure Controls
Framework (SCF)}. \url{https://securecontrolsframework.com}.

{[}47{]} R. Patir, K. Guo, S. Barua, A. Pathak, D. Gudimetla, J. Guo, H.
Hu, H. Cai. \emph{DualGauge: Automated Joint Security-Functionality
Benchmarking of Specification-Only Code Generation by LLMs and Coding
Agents}. arXiv:2511.20709v2, 2026.

{[}48{]} M. Vero, N. Mündler, V. Chibotaru, V. Raychev, M. Baader, N.
Jovanović, J. He, M. Vechev. ``BaxBench: Can LLMs Generate Correct and
Secure Backends?'' In \emph{Proc. 42nd International Conference on
Machine Learning (ICML 2025)}, PMLR 267:61344--61390, 2025.
arXiv:2502.11844.

{[}49{]} OWASP. \emph{OpenCRE: Open Common Requirement Enumeration}.
\url{https://opencre.org}; source code at github.com/OWASP/OpenCRE
(accessed 6 November 2026).

{[}50{]} N. Keller, S. Quinn, K. Scarfone, M. Smith, V. Johnson.
\emph{National Online Informative References (OLIR) Program: Overview,
Benefits, and Use}. NIST IR 8278 Rev.~1, February 2024.
\url{doi:10.6028/NIST.IR.8278r1}.

{[}51{]} K. Scarfone, M. Souppaya, M. Fagan. \emph{Mapping Relationships
Between Documentary Standards, Regulations, Frameworks, and Guidelines:
Developing Cybersecurity and Privacy Concept Mappings}. NIST IR 8477,
February 2024. \url{doi:10.6028/NIST.IR.8477}.

{[}52{]} NIST. \emph{Open Security Controls Assessment Language
(OSCAL)}, version 1.2.3, August 2026.
\url{https://pages.nist.gov/OSCAL/}.

{[}53{]} Cloud Security Alliance. \emph{Cloud Controls Matrix and CAIQ
v4.1}, January 2026.

{[}54{]} ISO/IEC 27034-1:2011. \emph{Information technology --- Security
techniques --- Application security --- Part 1: Overview and concepts}.

{[}55{]} Z. Syed, A. Padia, T. Finin, L. Mathews, A. Joshi. ``UCO: A
Unified Cybersecurity Ontology''. In \emph{AAAI Workshop on Artificial
Intelligence for Cyber Security}, 2016.

{[}56{]} M. Iannacone et al.~``Developing an Ontology for Cyber Security
Knowledge Graphs''. In \emph{Proc. 10th Annual Cyber and Information
Security Research Conference (CISR '15)}, ACM, 2015.
\url{doi:10.1145/2746266.2746278}.

{[}57{]} P. E. Kaloroumakis, M. J. Smith. \emph{Toward a Knowledge Graph
of Cybersecurity Countermeasures}. The MITRE Corporation, 2021.

{[}58{]} A. Souag, R. Mazo, C. Salinesi, I. Comyn-Wattiau. ``Reusable
knowledge in security requirements engineering: a systematic mapping
study''. \emph{Requirements Engineering} 21(2):251--283, 2016.
\url{doi:10.1007/s00766-015-0220-8}.

{[}59{]} S. Fenz, A. Ekelhart. ``Formalizing information security
knowledge''. In \emph{Proc. ASIACCS 2009}, pp.~183--194.
\url{doi:10.1145/1533057.1533084}.

{[}60{]} D. Firesmith. ``Specifying Reusable Security Requirements''.
\emph{Journal of Object Technology} 3(1):61--75, 2004.
\url{doi:10.5381/jot.2004.3.1.c6}.

{[}61{]} M. C. Suárez-Figueroa, A. Gómez-Pérez, M. Fernández-López.
``The NeOn Methodology framework: A scenario-based methodology for
ontology development''. \emph{Applied Ontology} 10(2):107--145, 2015.
\url{doi:10.3233/AO-150145}.

{[}62{]} M. Poveda-Villalón, A. Fernández-Izquierdo, M. Fernández-López,
R. García-Castro. ``LOT: An industrial oriented ontology engineering
framework''. \emph{Engineering Applications of Artificial Intelligence}
111:104755, 2022. \url{doi:10.1016/j.engappai.2022.104755}.

{[}63{]} A. R. Hevner, S. T. March, J. Park, S. Ram. ``Design Science in
Information Systems Research''. \emph{MIS Quarterly} 28(1):75--105,
2004. \url{doi:10.2307/25148625}.

\subsection{Appendix A. Sources of the
corpus}\label{appendix-a.-sources-of-the-corpus}

Table A1 summarises the 31 sources by group. Substrate items are the
source items of the similarity layer, curated rows are the rows of the
curated mapping file, and mapped rows are those with status mapped,
mapped by topic or inherited through a cross-reference. Objectives and
domains count the distinct objectives used as primary or secondary
anchor by the mapped rows. The scripts for ASVS and for the MCP security
documentation wrote \emph{direct} where the others wrote \emph{strong}.

\begin{landscape}\scriptsize\renewcommand{\arraystretch}{1.0}\setlength{\tabcolsep}{3pt}

\textbf{Table A1.} The 31 sources.

\begin{longtable}[]{@{}
  >{\raggedright\arraybackslash}p{(\linewidth - 18\tabcolsep) * \real{0.0533}}
  >{\raggedright\arraybackslash}p{(\linewidth - 18\tabcolsep) * \real{0.2533}}
  >{\raggedleft\arraybackslash}p{(\linewidth - 18\tabcolsep) * \real{0.0667}}
  >{\raggedleft\arraybackslash}p{(\linewidth - 18\tabcolsep) * \real{0.0667}}
  >{\raggedleft\arraybackslash}p{(\linewidth - 18\tabcolsep) * \real{0.0667}}
  >{\raggedleft\arraybackslash}p{(\linewidth - 18\tabcolsep) * \real{0.0533}}
  >{\raggedleft\arraybackslash}p{(\linewidth - 18\tabcolsep) * \real{0.0667}}
  >{\raggedright\arraybackslash}p{(\linewidth - 18\tabcolsep) * \real{0.2267}}
  >{\raggedleft\arraybackslash}p{(\linewidth - 18\tabcolsep) * \real{0.0800}}
  >{\raggedleft\arraybackslash}p{(\linewidth - 18\tabcolsep) * \real{0.0667}}@{}}
\toprule\noalign{}
\begin{minipage}[b]{\linewidth}\raggedright
Group
\end{minipage} & \begin{minipage}[b]{\linewidth}\raggedright
Source
\end{minipage} & \begin{minipage}[b]{\linewidth}\raggedleft
Substrate items
\end{minipage} & \begin{minipage}[b]{\linewidth}\raggedleft
Curated rows
\end{minipage} & \begin{minipage}[b]{\linewidth}\raggedleft
Mapped
\end{minipage} & \begin{minipage}[b]{\linewidth}\raggedleft
Out of scope
\end{minipage} & \begin{minipage}[b]{\linewidth}\raggedleft
Context only
\end{minipage} & \begin{minipage}[b]{\linewidth}\raggedright
Strength of mapped rows
\end{minipage} & \begin{minipage}[b]{\linewidth}\raggedleft
Objectives
\end{minipage} & \begin{minipage}[b]{\linewidth}\raggedleft
Domains
\end{minipage} \\
\midrule\noalign{}
\endhead
\bottomrule\noalign{}
\endlastfoot
A & OWASP ASVS 5.0.0 & 345 & 443 & 289 & 56 & 98 & direct 206, bounded
71, strong 12 & 33 & 9 \\
A & MITRE CAPEC 3.9 & 559 & 559 & 532 & 27 & 0 & strong 532 & 21 & 8 \\
A & CIS Controls 8.1.2 & 166 & 171 & 116 & 55 & 0 & strong 63, bounded
53 & 41 & 10 \\
A & MITRE CWE 4.19.1 (software development view) & 399 & 400 & 143 & 256
& 1 & bounded 77, strong 66 & 33 & 9 \\
A & ENISA multilayer framework for AI 2023 & 5 & 6 & 4 & 0 & 2 & bounded
4 & 3 & 3 \\
A & MCP security documentation 2025 & 13 & 14 & 12 & 1 & 1 & strong 6,
direct 4, bounded 2 & 5 & 3 \\
A & NIST SP 800-53 Rev.~5 & 1196 & 1196 & 869 & 327 & 0 & strong 674,
bounded 195 & 58 & 10 \\
A & OWASP DSOMM & 193 & 193 & 193 & 0 & 0 & strong 125, bounded 68 & 53
& 10 \\
A & OWASP Secure MCP Server Dev. 1.0 & 10 & 11 & 7 & 3 & 1 & bounded 5,
strong 2 & 7 & 6 \\
A & OWASP Third-Party MCP Servers 1.0 & 8 & 9 & 7 & 0 & 2 & strong 6,
bounded 1 & 6 & 4 \\
A & OWASP MCP Top 10 (beta) & 10 & 11 & 10 & 0 & 1 & strong 9, bounded 1
& 11 & 7 \\
A & OWASP Proactive Controls 2018 & 10 & 10 & 10 & 0 & 0 & strong 10 &
18 & 7 \\
A & OWASP SAMM 2.1 & 90 & 189 & 89 & 64 & 36 & strong 48, bounded 41 &
32 & 7 \\
A & OWASP Top 10:2021 & 10 & 10 & 10 & 0 & 0 & strong 10 & 23 & 9 \\
A & PCI Secure SLC 1.1 & 28 & 28 & 28 & 0 & 0 & strong 28 & 17 & 7 \\
A & SAFECode Agile 2012 & 29 & 29 & 29 & 0 & 0 & strong 29 & 12 & 5 \\
A & SAFECode Fundamental Practices 2018 & 17 & 17 & 17 & 0 & 0 & strong
17 & 18 & 9 \\
A & SAFECode Software Integrity Controls 2010 & 12 & 12 & 12 & 0 & 0 &
strong 12 & 10 & 5 \\
A & SLSA 1.0 Build Track & 14 & 14 & 14 & 0 & 0 & strong 13, bounded 1 &
7 & 3 \\
A & NIST SSDF 1.1 & 61 & 67 & 67 & 0 & 0 & strong 38, bounded 27, none 2
& 36 & 10 \\
B & EU CRA & 9 & 9 & 4 & 5 & 0 & bounded 4 & 4 & 4 \\
B & EU DORA & 24 & 24 & 23 & 1 & 0 & bounded 23 & 10 & 8 \\
B & EU NIS2 & 4 & 4 & 4 & 0 & 0 & bounded 4 & 4 & 4 \\
B & EU GDPR & 6 & 6 & 6 & 0 & 0 & bounded 6 & 4 & 4 \\
B & PCI Data Security Standard 4.0.1 & 217 & 217 & 205 & 12 & 0 &
bounded 205 & 47 & 9 \\
B & HIPAA Security Rule & 22 & 22 & 15 & 7 & 0 & bounded 15 & 10 & 8 \\
C & MITRE ATLAS & 278 & -- & -- & -- & -- & -- & -- & -- \\
C & OWASP LLM Top 10 2025 & 10 & -- & -- & -- & -- & -- & -- & -- \\
C & OWASP ML Top 10 (v0.3) & 10 & -- & -- & -- & -- & -- & -- & -- \\
C & NIST AI 100-2 E2025 & 53 & -- & -- & -- & -- & -- & -- & -- \\
C & NIST AI RMF 1.0 & 53 & -- & -- & -- & -- & -- & -- & -- \\
\end{longtable}

\end{landscape}

\end{document}